%% file: main.tex
\documentclass[sigconf, nonacm]{acmart}
\renewcommand\footnotetextcopyrightpermission[1]{}

\AtBeginDocument{%
  }

\setcopyright{acmlicensed}
\copyrightyear{2026}
\acmYear{2026}
\acmDOI{XXXXXXX.XXXXXXX}
\acmConference[Conference acronym 'XX]{Make sure to enter the correct
  conference title from your rights confirmation email}{August,
  2027}{Athens, Greece}
\acmISBN{978-1-4503-XXXX-X/2018/06}

\usepackage{xspace}
\usepackage[linesnumbered,ruled,vlined]{algorithm2e}
\usepackage{amsmath}
\usepackage{enumitem}
\usepackage{subcaption}
\usepackage{comment}
\usepackage{booktabs}
\usepackage{tikz}
\usepackage{multirow}
\usepackage{pifont}
\usepackage{makecell}

\usepackage[table,x11names]{xcolor}

\newcommand{\squishlist}{
	\begin{list}{$\bullet$}
		{ \setlength{\itemsep}{1pt}
			\setlength{\parsep}{1pt}
			\setlength{\topsep}{2.5pt}
			\setlength{\partopsep}{0.5pt}
			\setlength{\leftmargin}{1em}
			\setlength{\labelwidth}{1em}
			\setlength{\labelsep}{0.6em}
		}
	}
	\newcommand{\squishend}{
	\end{list}
}
            
\newcommand\frbox[2][draw=black!30]{%
    \tikz[baseline]\node[%
        inner ysep=0pt, 
        inner xsep=2pt, 
        anchor=text, 
        rectangle, 
        rounded corners=1mm,
        #1] {\strut#2};%
}

\newcommand{\circled}[1]{\raisebox{.5pt}{\textcircled{\raisebox{-.9pt} {#1}}}}

\definecolor{darkyellow}{RGB}{204, 153, 0}
\definecolor{darkgreen}{RGB}{0, 128, 0}
\definecolor{darkpurple}{RGB}{128, 0, 128}

\newcommand{\rone}[1]{{#1}}
\newcommand{\rtwo}[1]{{#1}}
\newcommand{\rthree}[1]{{#1}}
\newcommand{\rmeta}[1]{{#1}}

\DeclareMathOperator*{\argmax}{arg\,max}

\theoremstyle{theorem}
\newtheorem{definition}{Definition}

\newcommand{\ignore}[1]{}

\newcommand{\name}{\textsc{Chimera}\xspace}

\newcommand{\stitle}[1]{\vspace*{0.4em}\noindent{\bf #1.\/}}

\begin{document}

\title{\name: Efficient Multi-Vector Retrieval via GPU--CPU Co-Processing}

\author{Yanqi Chen}
\affiliation{%
  \institution{Univ.\ of Massachusetts}
  \city{Amherst}
  \country{USA}
}
\email{yanqichen@cs.umass.edu}

\author{Juelin Liu}
\affiliation{%
  \institution{Univ.\ of Massachusetts}
  \city{Amherst}
  \country{USA}
}
\email{juelinliu@cs.umass.edu}

\author{Alexandra Meliou}
\affiliation{%
  \institution{Univ.\ of Massachusetts}
  \city{Amherst}
  \country{USA}
}
\email{ameli@cs.umass.edu}

\author{Xiao Yan}
\affiliation{%
 \institution{Wuhan University}
 \city{Wuhan}
 \country{China}
}
\email{yanxiaosunny@whu.edu.cn}



\begin{abstract}
  Multi-vector retrieval has become an important primitive for fine-grained matching in information retrieval, with emerging applications in areas such as recommender systems and bioinformatics. However, its high computational complexity and memory costs make low-latency retrieval difficult. Prior systems have attempted to optimize query latency, but their designs remain CPU-centric. While GPUs offer substantial computational advantages, their limited memory capacity necessitates a heterogeneous architecture in which the dataset resides in host memory and the GPU serves as an accelerator. The state-of-the-art GPU-based system, PLAID, is bottlenecked by CPU--GPU data movement, as vector data must be transferred from host memory to the GPU at query time. We propose \name, a GPU--CPU co-processing system for multi-vector retrieval that eliminates this transfer bottleneck. \name stores highly compressed, low-precision quantization codes on the GPU while maintaining high-precision data in CPU memory. At query time, it leverages GPU-resident data for efficient candidate generation and pruning, and further refines results through a CPU-GPU collaborative scoring scheme that completely avoids vector data transfer while enabling computation overlap. Experiments on real-world datasets demonstrate that \name significantly outperforms existing approaches, achieving up to $59.5\times$ higher QPS at the same recall level. \rtwo{Our code is available at https://github.com/iidyc/Chimera.}
\end{abstract}





\maketitle

\input{sections/intro}

\input{sections/prelim}
\input{sections/method}
\input{sections/impl-details}
\input{sections/experiments}
\input{sections/related-work}
\input{sections/conclusion}


\bibliographystyle{ACM-Reference-Format}
\bibliography{references}

\end{document}

%% file: sections/intro.tex
\section{Introduction}\label{sec:intro}

The rapid proliferation of deep learning systems has made representation learning a central component of modern data processing pipelines. In particular, embedding techniques have proven to be a powerful tool for capturing the semantic structure of data objects, enabling meaningful comparisons in a continuous vector space across domains such as text, images, and graphs.

Traditional approaches often rely on representing each object 
with a single dense vector, and a similarity-search query retrieves the top-$k$ most similar vectors to identify its related objects. This paradigm has been highly successful, especially when combined with approximate nearest neighbor search (ANNS) techniques, powered by specialized vector indexes such as IVF~\cite{pq} and proximity graph indexes~\cite{cagra}, which enable efficient retrieval in large-scale datasets while preserving meaningful semantic relationships. Single-vector embeddings remain widely used across applications in 
data mining~\cite{huang2017query,iwasaki2016pruned}, pattern recognition~\cite{pattern,object}, recommender systems~\cite{mind}, and retrieval augmented generation (RAG) for LLMs~\cite{gao2024retrievalaugmented}, owing to their simplicity and strong empirical performance.


\begin{figure}[!t]
  \centering
  \includegraphics[width=\linewidth]{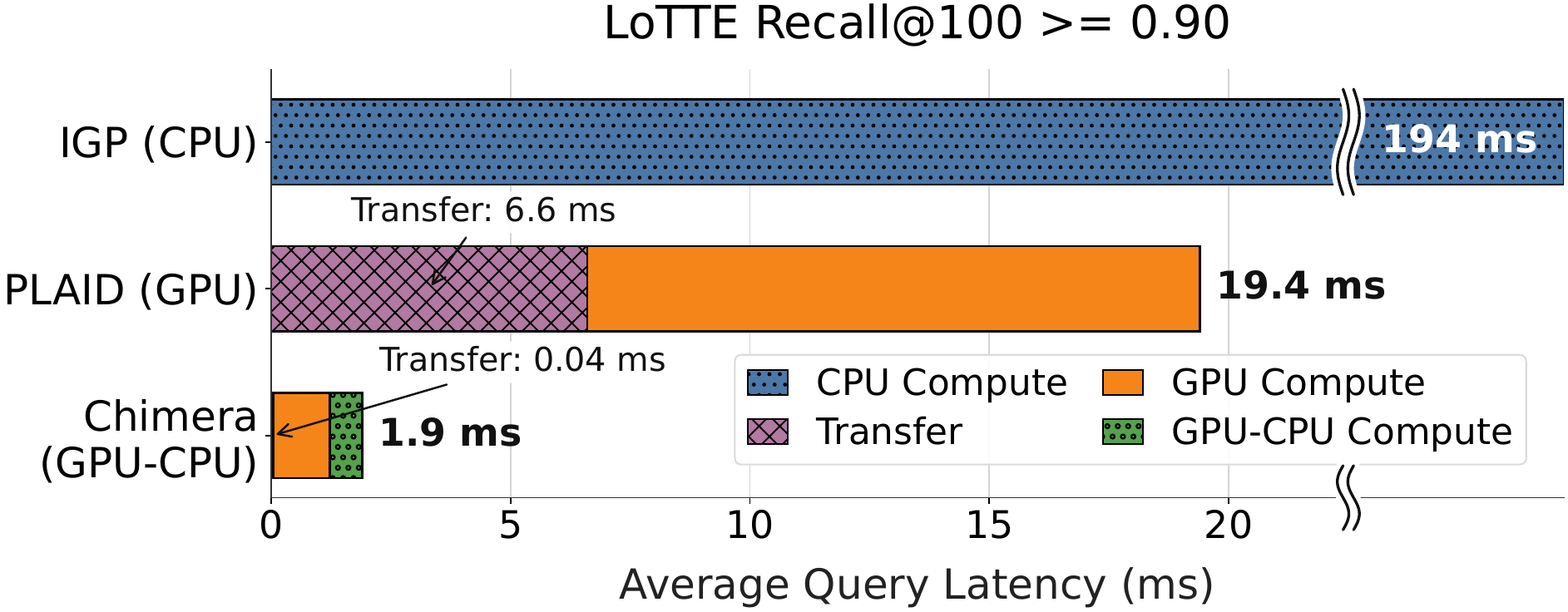}
  \vspace{-6mm}
  \caption{
  Multi-vector retrieval latency for IGP, PLAID, and our \name on the LoTTE dataset. IGP runs on CPU, PLAID uses GPU, and \name uses both GPU and CPU. 
  }
  \label{fig:lotte-plaid-chimera-breakdown}
\end{figure}

\stitle{Multi-vector retrieval} \looseness-1
However, representing each object with a single vector may not adequately capture the rich semantics of complex data (e.g., different topics in a  document or different components of an image). Multi-vector representations, where each data object is modeled as a set of vectors, have emerged as a promising alternative~\cite{colbert}. 
By allowing multiple embeddings to jointly characterize different aspects, components, or modalities of an object, multi-vector models can better preserve fine-grained semantic information, improve expressiveness, and enable more flexible matching mechanisms compared to single-vector representations.



\rmeta{Multi-vector representations arise in several domains, and we present three examples here. ColBERT-style information retrieval is the primary setting targeted by \name; we discuss recommender systems and bioinformatics as broader, emerging uses of multi-vector representations:}

\begin{description}[topsep=2pt, leftmargin=6pt, labelsep=0pt]
    \item \frbox{\emph{Information retrieval}~\cite{colbert}:} 
    The late-interaction mechanism introduced by ColBERT for document retrieval represents documents and queries as sets of token-level vectors\footnote{To ease discussion, we follow information retrieval terminology, referring to multi-vectors as \textit{documents} and their constituent single-vectors as \textit{tokens}.} 
    %
    rather than single dense embeddings. This design enables fine-grained similarity matching, preserving rich semantic signals at the token level and thus leading to more accurate and robust retrieval performance.
    
    \item \frbox{\emph{Recommender systems}~\cite{mind}:} 
    User interaction histories often encompass multiple distinct interests, which cannot be captured by a single embedding. Multi-interest recommendation models represent a user as a set of vectors, each encoding a specific interest, enabling the candidate items to be matched more precisely across diverse user preferences.
    
    \item \frbox{\emph{Bioinformatics}~\cite{bioinfo}:} 
    Proteins exhibit complex structures that are not well captured by a single embedding. Representing each protein as a set of vectors corresponding to different substructures enables fine-grained matching and improves tasks such as substructure search and remote homology detection.
\end{description}

\stitle{Challenges} 
Unfortunately, the transition from single-vector point-to-point comparisons to set-to-set distance evaluations introduces two profound challenges:
 
(1)~Multi-vector similarity functions are \emph{inherently computation-intensive}. Unlike single-vector retrieval, where a single dot product or distance evaluation suffices, multi-vector retrieval requires aggregating pairwise similarities across two sets of vectors. This leads to substantially higher computational complexity, making it difficult to meet the stringent latency requirements of real-time applications without hardware acceleration.  Figure~\ref{fig:lotte-plaid-chimera-breakdown} profiles IGP~\cite{igp}, the state-of-the-art CPU-based multi-vector retrieval engine, showing a query latency close to 200ms, while the latency of single vector search is usually below 10ms when all vectors fit in memory.

(2)~With multi-vector representations, the \emph{dataset size increases dramatically}. In many practical systems, each data object is represented by tens to hundreds of vectors; this expansion inflates the storage footprint by up to two orders of magnitude compared to single-vector datasets, placing significant pressure on memory capacity and data movement, and hindering the potential gains of hardware acceleration through GPUs when datasets do not fit in GPU memory.  Our profiling in Figure~\ref{fig:lotte-plaid-chimera-breakdown} shows that PLAID~\cite{plaid}, the state-of-the-art GPU-based multi-vector retrieval engine, incurs a severe data movement bottleneck, with data transfer accounting for up to 34\% of total execution time. The LoTTE dataset, used in this experiment, contains 2.4M documents and takes up 1.2GB if each document is represented by a single vector. Using multi-vector representation, the LoTTE dataset takes up about 130GB, while an NVIDIA V100 GPU has only 32GB memory. \rthree{Although newer GPUs provide larger HBM capacity, retrieval corpora are also continuously growing; therefore, the relevant factor is the ratio between dataset size and device memory rather than their absolute sizes. Newer hardware shifts the point at which a dataset exceeds HBM, but does not eliminate this out-of-memory regime.}

\stitle{Shortcomings of prior art}
These challenges jointly motivate the need for a CPU-GPU co-processing architecture. While GPUs are well-suited for accelerating the massive number of vector similarity computations, their limited memory capacity 
prevents them from hosting the entire multi-vector dataset. As a result, the dataset must reside in main memory, with the GPU acting as a compute accelerator for selected portions of the workload. This setting is particularly relevant in cost-sensitive or resource-constrained deployments, where high-end GPUs with large memory are not available.

However, existing systems struggle to fully exploit such a heterogeneous architecture. Although several systems (e.g., EMVB~\cite{emvb}, WARP~\cite{warp}, DESSERT~\cite{dessert}, MUVERA~\cite{muvera} and IGP~\cite{igp}) have been proposed to improve the efficiency of multi-vector retrieval, they remain largely CPU-centric. To the best of our knowledge, PLAID~\cite{plaid} is the only prior system that provides a GPU-based implementation for multi-vector retrieval. However, as we saw in Figure~\ref{fig:lotte-plaid-chimera-breakdown}, PLAID's performance remains suboptimal.  This is due to three reasons: 
(1)~PLAID's on-demand transfer of vectors to GPU at query time causes severe data movement bottleneck due the low arithmetic intensity of multi-vector scoring: each inner product over $d$-dimensional embeddings requires $2d$ FLOPs but transfers $4d$ bytes, yielding only $0.5$ FLOPs/byte. 
\rthree{Although batching can increase the effective arithmetic intensity to approximately 20 FLOPs/byte, this remains far below the compute-to-PCIe balance of modern GPUs, which ranges from approximately 300 to 470 FLOPs/byte across V100, A100 and H100 platforms~\cite{nvidia_hopper_architecture_blog}.}
Given the large gap between GPU compute throughput and PCIe bandwidth, streaming document tokens from host memory makes execution fundamentally bandwidth-bound, leaving the GPU underutilized.\rthree{\footnote{\rthree{Although newer platforms such as NVIDIA Grace Hopper (GH200) provide a unified CPU--GPU memory space that can reduce explicit data movement, such systems remain substantially more expensive and less widely available than conventional PCIe-based GPUs. Therefore, a system that performs well on lower-cost, widely available commodity GPUs such as V100 remains practically valuable, especially for users who cannot rely on abundant access to the newest hardware generation.}}}
(2)~PLAID uses a two-stage retrieval pipeline based on an inverted file index (IVF).  Because IVF partitions the embedding space coarsely, the resulting clusters often have low selectivity, failing to filter many irrelevant documents. Moreover, centroid scores provide only a rough approximation of true document similarity, which further reduces filtering quality and degrades overall efficiency.
(3)~PLAID implements its GPU computation using high-level PyTorch primitives rather than custom-designed kernels tailored to multi-vector retrieval. While this choice simplifies development, it limits performance because these general-purpose operators are not optimized for the specific access patterns and fine-grained parallelism required by multi-vector scoring.

\stitle{Our solution} 
We propose \name, a GPU–CPU co-processing system for multi-vector retrieval that eliminates the data transfer bottleneck by ensuring the GPU never fetches document data from main memory at query time. Rather than treating the GPU as an on-demand accelerator, \name assigns complementary roles to the two processors \rthree{through a combination of algorithmic contributions, system-level execution, and algorithm--system co-design.} 

\rthree{
At the algorithmic level, \name introduces a multi-stage retrieval pipeline that combines aggressive low-precision filtering with accurate high-precision refinement. During indexing, \name compresses the dataset using the state-of-the-art RaBitQ quantization algorithm~\cite{erabitq}: compact 1-bit codes are stored on the GPU for fast approximate scoring, while higher-precision quantization codes are retained in CPU memory for accurate refinement. In contrast with PLAID's large clusters, \name partitions vectors into fine-grained clusters and builds a proximity graph over their centers, enabling efficient search over a larger number of more selective partitions. The resulting index, together with the low-precision representations, allows \name to prune the search space aggressively before invoking expensive full-precision scoring.

At the system level, \name places the 1-bit codes and auxiliary indexes entirely in GPU memory, so the only per-query data transferred to the GPU is the query multi-vector. The CPU stores the higher-precision codes and scores only the remaining candidates. To maximize processor utilization, \name pipelines CPU scoring of earlier candidate chunks with GPU processing of later chunks, enabling concurrent execution without transferring document representations over PCIe.

These algorithmic and system choices are jointly designed around the heterogeneous memory hierarchy. The low-precision filtering stages allow for effective pruning, and their compact representations fit in GPU memory and can be evaluated efficiently in parallel; the more accurate refinement stage is assigned to the CPU, where substantially larger memory capacity is available. \name further realizes this design with customized GPU kernels for its key operations: LUT-based binary inner-product kernels replace bit-wise conditional accumulation with table lookups to eliminate branch divergence, while stage-specific score aggregation kernels reduce atomic contention during candidate refinement and improve memory coalescing during document scoring.

Together, these contributions exploit the complementary strengths of the two processors---memory capacity on the CPU and parallelism on the GPU---while avoiding query-time document transfer, thereby achieving both high recall and high efficiency.
}

\stitle{Contributions}  Our work makes the following contributions.
\begin{itemize}[topsep=0pt, leftmargin=6pt]
    \item We identify the key challenges in multi-vector retrieval---namely excessive data volume and intensive computation---and motivate the need for a hybrid GPU--CPU system. [Sections~\ref{sec:intro}--\ref{sec:prelim}]
    \item We propose \name, a novel GPU--CPU co-processing system for multi-vector retrieval, that tailors its index structure and data layout for this hybrid setting. Through the core designs of hybrid-precision storage and a concurrent GPU--CPU collaborative scoring mechanism, \name eliminates data transfer bottlenecks while fully utilizing available compute resources. [Section~\ref{sec:methodology}]
    \item We introduce hardware-aware optimizations, including look-up table (LUT)-based binary inner product kernels and stage-specific document score aggregation kernels, to accelerate multi-vector similarity computation on GPUs. [Section~\ref{sec:sys-opt}]
    \item We conduct comprehensive evaluations on three large-scale, real-world multi-vector datasets. Experimental results demonstrate that our system offers clear and consistent improvements, achieving up to 16.0$\times$ higher query throughput compared to the prior state of the art at the same recall level. [Section~\ref{sec:expr}]
\end{itemize}

%% file: sections/prelim.tex
\section{Background and preliminaries}\label{sec:prelim}

In this section, we formalize our problem setting, and review important tools from prior work that \name employs in a novel way in its design.  Specifically, we use RaBitQ~\cite{rabitq, erabitq} for vector compression (Section~\ref{sec:quant}), and IVF and proximity graph indexes to efficiently retrieve relevant documents (Section~\ref{sec:indexes}).  After summarizing these techniques, we discuss how they are incorporated in \name's design in Section~\ref{sec:methodology}.

\subsection{The multi-vector retrieval setting}
We first define the similarity function between two multi-vectors using Chamfer similarity. We then formalize the multi-vector retrieval (MVR) problem based on this similarity. Finally, we introduce Recall@$k$ as the evaluation metric for retrieval effectiveness.

\begin{definition}[Chamfer Similarity]
Given two sets of vectors $Q=\{q_1,q_2,\dots,q_m\}$ and $V=\{v_1,v_2,\dots,v_n\}$ in $\mathbb{R}^d$, the Chamfer similarity 
is defined as
$ \mathcal{F}(Q,V) = \sum_{q \in Q} \max_{v \in V} \langle q, v \rangle$,
where $\langle q, v \rangle$ denotes the inner product between vectors $q$ and $v$.
\end{definition}


We note that Chamfer similarity has a long history in the machine learning literature for comparing sets of embeddings~\cite{sudderth2004visual,bakshi10near,WanCLYZY019}, and is the default similarity metric in existing MVR systems, including PLAID~\cite{plaid}, EMVB~\cite{emvb}, and IGP~\cite{igp}, among others.

\begin{definition}[Multi-Vector Retrieval (MVR)]
Let $\mathcal{D}$ be a dataset of multi-vectors. Given a query multi-vector $Q$ and an integer $k$, the multi-vector retrieval problem aims to return a set $\mathcal{R} \subset \mathcal{D}$ of top-$k$ similar multi-vectors to $Q$ with respect to Chamfer similarity:
%
\begin{equation}
    \mathcal{R} = \argmax_{\substack{\mathcal{R} \subset \mathcal{D} \\ |\mathcal{R}| = k}} \sum_{V \in \mathcal{R}} \mathcal{F}(Q, V)
\end{equation}
\end{definition}

In practice, computing exact Chamfer similarity---known to admit no sub-quadratic time algorithm~\cite{bakshi10near}---across all documents in a large-scale dataset is computationally prohibitive and infeasible with strict latency requirements. Thus, retrieval algorithms typically return approximate solutions.
We adopt the standard definition of Recall@$k$ as the primary metric of retrieval effectiveness.

\begin{definition}[Recall@$k$]
Let $\mathcal{R}$ denote the ground-truth top-$k$ multi-vectors for a query $Q$, and let $\mathcal{R}'$ denote the set of top-$k$ multi-vectors returned by an algorithm. 
Then, $\mathrm{Recall@}k = \frac{|\mathcal{R} \cap \mathcal{R}'|}{k}$.
\end{definition}

\subsection{RaBitQ for vector quantization}\label{sec:quant}

RaBitQ~\cite{rabitq, erabitq} is a quantization technique for vector compression.\footnote{In this paper, RaBitQ refers to the extended-bits version~\cite{erabitq}.} 
Specifically, it compresses each dimension of the vector using $B$ bits. It provides an unbiased estimator for inner product between quantized vectors.

\stitle{Vector quantization} Given a data vector $\mathbf{o}_r$ and a query vector $\mathbf{q}_r$, RaBitQ first normalizes them into unit vectors:
\begin{equation}
    \mathbf{o}:=\frac{\mathbf{o}_r}{\lVert \mathbf{o}_r \rVert},\ \mathbf{q}:=\frac{\mathbf{q}_r}{\lVert \mathbf{q}_r \rVert}.
\end{equation}
\noindent Then the inner product between the original vectors (i.e., $\mathbf{o}_r$ and $\mathbf{q}_r$) can be expressed as
\begin{equation}
    \langle \mathbf{q}_r,\mathbf{o}_r \rangle=\lVert \mathbf{q}_r \rVert \lVert \mathbf{o}_r \rVert \langle \mathbf{q},\mathbf{o} \rangle,
\end{equation}
\noindent where $\lVert \mathbf{o}_r \rVert$ can be precomputed and stored offline and $\lVert \mathbf{q}_r \rVert$ can be computed once and reused during retrieval. As such, RaBitQ quantizes the normalized data vector $\mathbf{o}$ and focuses on estimating the inner product $\langle \mathbf{q}, \mathbf{o} \rangle$.

To conduct quantization, RaBitQ rotates the normalized data vector $\mathbf{o}$ using a random orthonormal matrix $P$ to obtain $\mathbf{o}_p:=P\cdot\mathbf{o}$. Since $P$ is orthonormal, the inner product is preserved after random rotation. As such, the goal of RaBitQ is to compress $\mathbf{o}_p$ while providing an accurate estimation of $\langle \mathbf{q}_p,\mathbf{o}_p \rangle$. 
Specifically, RaBitQ quantizes $\mathbf{o}_p$ using the following codebook $\mathcal{G}_r$:
\begin{equation}
\begin{gathered}
    \mathcal{G}:=\left\{-\frac{2^B-1}{2}+u\ \Bigg{|}\ u=0,1,2,\dots,2^B-1 \right\}^D\\
    \mathcal{G}_r:=\left\{\frac{\mathbf{y}}{\lVert\mathbf{y}\rVert}\Bigg{|}\ \mathbf{y}\in\mathcal{G} \right\}^D.
\end{gathered}
\end{equation}

A data vector $\mathbf{o}_p$ finds its nearest codeword in $\mathcal{G}_r$ as its quantized vector $\mathbf{\bar{o}}_p$ and stores the corresponding quantization code $\mathbf{\bar{o}}_b\in\{0,1,2,\dots,2^B-1\}^D$.

\stitle{Inner product estimation} Given the quantized vector $\mathbf{\bar{o}}_p$ and the rotated normalized query vector $\mathbf{q}_p$, RaBitQ estimates $\langle \mathbf{q}_p,\mathbf{o}_p \rangle$ as
\begin{equation}
    \langle \mathbf{q}_p,\mathbf{o}_p \rangle\approx\frac{\langle \mathbf{q}_p,\mathbf{\bar{o}}_p \rangle}{\langle \mathbf{o}_p,\mathbf{\bar{o}}_p \rangle}
\end{equation}
\noindent where $\langle \mathbf{o}_p,\mathbf{\bar{o}}_p \rangle$ can be computed offline and stored. Thus, we only need to compute $\langle \mathbf{q}_p,\mathbf{\bar{o}}_p \rangle$. Is has been shown that the estimator is unbiased and has empirically tight error bound~\cite{erabitq}. 

Let $\mathbf{\bar{y}}$ be the corresponding vector of the quantized vector $\mathbf{\bar{o}}$ in $\mathcal{G}$, i.e., $\mathbf{\bar{o}}=\mathbf{\bar{y}}/\lVert \mathbf{\bar{y}}\rVert$. Note that the quantization code $\mathbf{\bar{o}}_b$ and $\mathbf{\bar{y}}$ has relationship $\mathbf{\bar{y}}=\mathbf{\bar{o}}_b-\frac{2^B-1}{2}\cdot\mathbf{1}_D$, where $\mathbf{1}_D$ is the $D$-dimensional vector whose coordinates are all ones. Then $\langle \mathbf{q}_p,\mathbf{\bar{o}}_p \rangle$ can be computed as
\begin{equation}
    \langle \mathbf{q}_p,\mathbf{\bar{o}}_p \rangle=\frac{1}{\lVert \mathbf{\bar{y}}\rVert}\left(\langle \mathbf{q}_p,\mathbf{\bar{o}}_b \rangle-\frac{2^B-1}{2}\sum_{i=1}^D\mathbf{q}_p[i]\right)
\end{equation}

\subsection{Indexes for single-vector search}\label{sec:indexes}

\name's engine relies on two types of single-vector indexes to identify relevant document tokens for each query token, in order to limit the number of distance computations. We briefly discuss the indexes used in \name below.

\begin{figure}[!t]
	\centering
	\includegraphics[width=0.4\textwidth]{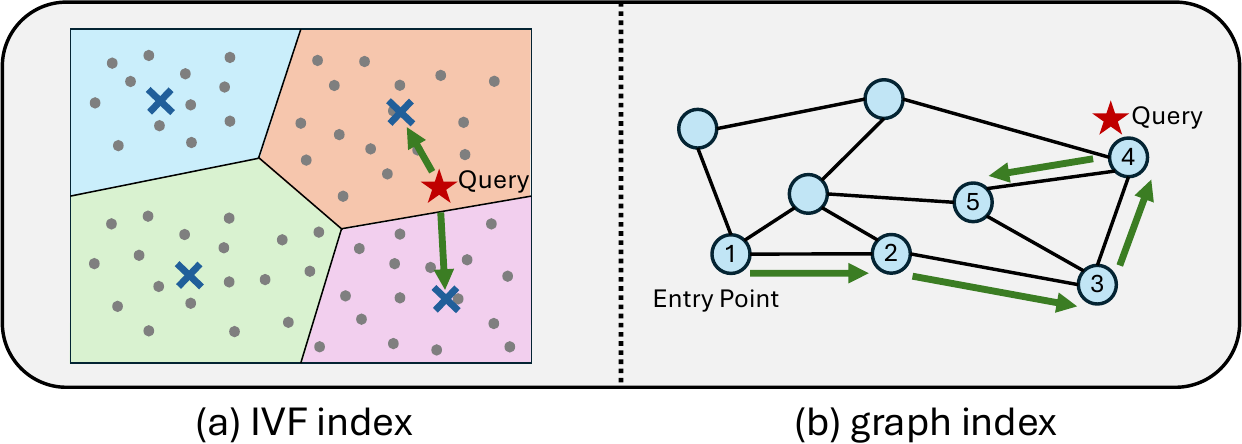}
    \vspace{-4mm}
	\caption{
    Vector indexes for single-vector search. 
    (a)~IVF index partitions the vector space into clusters and restricts search to a small subset of relevant clusters. (b)~Proximity graph connects nearby vectors and performs search via graph traversal over local neighborhoods.
    }
	\label{fig:index}
	\Description{}
\end{figure}

\stitle{Inverted file} The inverted file (IVF) index is a widely used structure for accelerating single-vector search by partitioning the data into coarse clusters and restricting search to only a small subset of them. It is typically constructed by learning a set of centroids (e.g., via k-means) and assigning each database vector to its nearest centroid, forming posting lists that store the vector ids assigned to each centroid. At query time, the query vector is first compared against all centroids to select the top-$n_{\text{probe}}$ closest clusters, and the search is then limited to vectors within these clusters, where distances or similarities are computed to retrieve the top-$k$ nearest neighbors, as illustrated in Figure~\ref{fig:index}(a).

\stitle{Proximity graph} Proximity graph is the state-of-the-art index for single-vector search as it requires significantly fewer distance computations to achieve the same recall as other indexes, such as IVF. The core idea of proximity graph is to organize vectors as nodes in a graph where edges connect nearby points, enabling efficient search via graph traversal, as illustrated in Figure~\ref{fig:index}(b). The graph is constructed by linking each node to a limited number of nearest neighbors, often with pruning or diversification strategies to ensure good navigability. At query time, the search starts from one or a few entry points and iteratively moves to neighboring nodes that are closer to the query until convergence. This approach efficiently explores local neighborhoods in the dataset and often achieves high recall with sublinear computation.

\stitle{Memory-efficiency tradeoff} While proximity graph indexes are generally more efficient, a key drawback is their substantial memory overhead, as the graph structure, which consists of explicit neighbor links for each vector, requires storage proportional to the dataset size. In contrast, IVF incurs much lower memory overhead: it only requires a relatively small set of centroids and lightweight posting lists of vector identifiers. Although the underlying vectors in graph-based indexes can be compressed using quantization techniques, the graph structure itself is inherently difficult to compress without degrading navigability and search performance. As a result, in memory-constrained settings, IVF-based approaches are often preferred as the base index, as they achieve a more favorable balance between memory usage and retrieval efficiency.
\name combines both index types in a way that leverages their respective advantages (discussed in Section~\ref{sec:methodology}).


%% file: sections/method.tex

\begin{figure*}[!t]
	\centering
	\includegraphics[width=\textwidth]{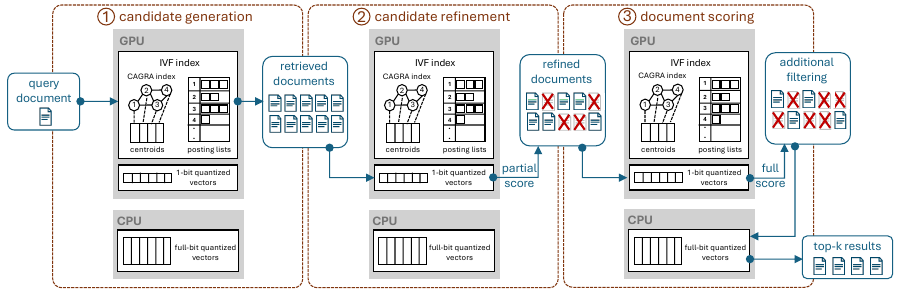}
    \vspace{-4mm}
	\caption{
    Overview of \name's query processing pipeline. \circled{1}~Candidate generation: query tokens probe IVF posting lists via CAGRA (a proximity graph index) to retrieve initial candidates on the GPU. \circled{2}~Candidate refinement: \name uses 1-bit codes to compute partial scores and prune candidates. \circled{3}~Document scoring: GPU computes the complete 1-bit document scores to aggressively filter the remaining candidates, and the CPU computes the full-bit scores to produce the final top-$k$ results.
    }
	\label{fig:overview}
	\Description{}
\end{figure*}

\section{\name: A hybrid co-processing architecture}
\label{sec:methodology}
Figure~\ref{fig:overview} illustrates the architecture and processing pipeline of \name. To balance efficiency and GPU memory constraints, \name adopts a heterogeneous design: compact $1$-bit quantization codes and IVF-related index data are resident on the GPU for high-throughput filtering, while full-bit codes are stored in CPU memory for accurate scoring. All GPU-resident data are preloaded before query processing, and only query token embeddings are transferred at runtime, avoiding per-query movement of large document data over PCIe. The system further overlaps GPU-based coarse scoring with CPU-based fine-grained evaluation via pipelined execution, maximizing utilization while minimizing data transfer overhead. This design enables high recall with substantially reduced computation and memory traffic.

\looseness-1
Specifically, given a query document, the system executes a three-stage retrieval pipeline consisting of \emph{candidate generation}, \emph{candidate refinement}, and \emph{document scoring}, as outlined in Algorithm~\ref{alg:overview}. 
In the first stage, each query token probes the IVF index to retrieve tokens from the posting lists of nearby clusters. To improve efficiency, the system first queries the proximity graph index built on the IVF centroids to identify the nearest centroids; it then fetches their corresponding posting lists, forming an initial set of candidate documents.
In the second stage, \name performs a lightweight refinement on the GPU using only the $1$-bit quantization codes, computing partial document scores based on retrieved tokens to prune the candidate set aggressively. 
In the final stage, the remaining documents are scored through a CPU–GPU collaborative process. The GPU computes scores using $1$-bit codes for coarse filtering, while the CPU leverages full-bit codes for higher accuracy, and their results are combined to produce the final top-$k$ documents.

\begin{algorithm}[!t]
	\caption{Overview of query processing in \name}
	\label{alg:overview}
    {\small
        \KwIn{Query $Q$, GPU resident data ($\mathsf{CAGRA}, \mathsf{IVF}, \mathsf{code}_{1\text{b}}$), CPU resident data ($\mathsf{code}_{\text{fb}}$), hyperparameters $n_{\mathsf{probe}}, k_{\mathsf{refine}}, k_{\mathsf{fb}}$}
        \KwOut{Top-$k$ documents}
        \BlankLine
        \tcc{Candidate Generation}
        $\mathsf{centroids} \leftarrow \mathsf{searchCAGRA}(Q, \mathsf{CAGRA}, n_{\mathsf{probe}})$\\
        $\{L_q\}_{q\in Q}\leftarrow \mathsf{fetchPostingList}(\mathsf{centroids},\mathsf{IVF})$\\
        $S\leftarrow \mathsf{findSourceDocuments}(\{L_q\}_{q\in Q})$ \\
        \tcc{Candidate Refinement}
        $\hat{F} \leftarrow \mathsf{partialScore}(Q, S, L_Q, \mathsf{code}_{\text{1b}})$\\
        $S_{\mathsf{refine}}\leftarrow \mathsf{topK}(\hat{F},S,k_{\mathsf{refine}})$\\
        \tcc{Document Scoring}
        $F,S_{\mathsf{fb}}\leftarrow \mathsf{collaborativeScoring}(S_{\mathsf{refine}},\mathsf{code}_{1\text{b}},\mathsf{code}_{\text{fb}},k_{\mathsf{fb}})$\\
        \Return $\mathsf{topK}(F,S_{\text{fb}},k)$
    }
\end{algorithm}

We proceed to describe \name's components and processing steps in more detail. Section~\ref{sec:offline} describes the data structures and compression that occur offline, and Section~\ref{sec:online} details the three processing steps of candidate generation, candidate refinement, and document scoring that occur during online retrieval.

\subsection{Offline index building} \label{sec:offline}

Akin to existing systems (e.g., PLAID~\cite{plaid}), \name constructs an IVF index on the document token embeddings. However, in contrast to PLAID, which uses relatively coarse partitions, \name adopts a much finer-grained clustering strategy by constructing a large number of IVF clusters. This design reduces the number of vectors per cluster, leading to more selective candidate generation and fewer irrelevant documents being carried into later stages. To efficiently search over this enlarged set of centroids, we build a CAGRA~\cite{cagra} index on the cluster centroids. This GPU-based proximity graph index enables fast identification of the top-$k$ nearest centroids for each query token, allowing \name to avoid expensive brute-force scans while maintaining high recall with significantly improved filtering quality.

To compress the dataset, we apply RaBitQ quantization on the document token embeddings. Specifically, each token embedding is encoded and stored as a quantized vector, where each dimension is represented using $B$ bits. In practice, we typically use $B=4$, which yields approximately $8\times$ compression while maintaining nearly identical recall compared to uncompressed vectors. The quantization code is stored in two parts: the 1-bit part (the most significant bits of the codes) and the full-bit codes. 


\subsection{Online query processing} \label{sec:online}
During retrieval, the IVF centroids, posting lists, and the CAGRA index are transferred to GPU memory. Because GPU memory is limited, only the $1$-bit portion of the RaBitQ codes is stored on the GPU, while the full-bit codes remain in main memory. 
\name takes as input a query document, the desired result size $k$, and a set of hyperparameters $(n_{\mathsf{probe}}, k_{\mathsf{refine}}, k_{\mathsf{full\text{-}bit}})$ that control the retrieval quality. The parameter $n_{\mathsf{probe}}$ determines how many IVF clusters are explored during candidate generation, $k_{\mathsf{refine}}$ sets the size of the candidate set retained after refinement, and $k_{\mathsf{full\text{-}bit}}$ specifies how many candidates are evaluated using full-precision scoring in the final stage. By progressively narrowing the search space across stages, these parameters enable an effective balance between efficiency and retrieval accuracy.

The retrieval pipeline of \name consists of three stages: \emph{candidate generation}, \emph{candidate refinement}, and \emph{document scoring}. The candidate generation and refinement stages are executed entirely on the GPU, without CPU–GPU data transfer except for uploading the query document. The document scoring stage involves both CPU and GPU computation, with dedicated coordination to maximize computation overlap while minimizing data transfer overhead.  We proceed to describe these three steps.


\stitle{\circled{1}~Candidate generation} The candidate generation (CG) stage queries the IVF index using the query tokens to identify a set of candidate documents that contain potential top-$k$ documents.  The key intuition is that documents containing tokens that are close to query tokens are more likely to be relevant.

Specifically, for each query token $q\in Q$, the top-$n_{\text{probe}}$ cluster centroids $C_q=\{c_1,c_2,\dots,c_{n_{\text{probe}}}\}$ ranked by distance to $q$ are retrieved using the CAGRA index. Each centroid $c_i$ corresponds to a cluster of document tokens $P_i$ (i.e., posting list). The candidate token list retrieved by each query token $q$ is thus
$$L_q=\bigcup_{i=1}^{n_{probe}}P_i.$$
The initial candidate document set is then constructed by taking the union of the source documents of all retrieved tokens.
$$S = \left\{V \mid \exists\ v \in V, \; v \in \bigcup_{q\in Q} L_q \right\}$$


\begin{figure*}[t]
    \centering
    \begin{subfigure}{0.48\linewidth}
        \centering
        \includegraphics[width=\linewidth]{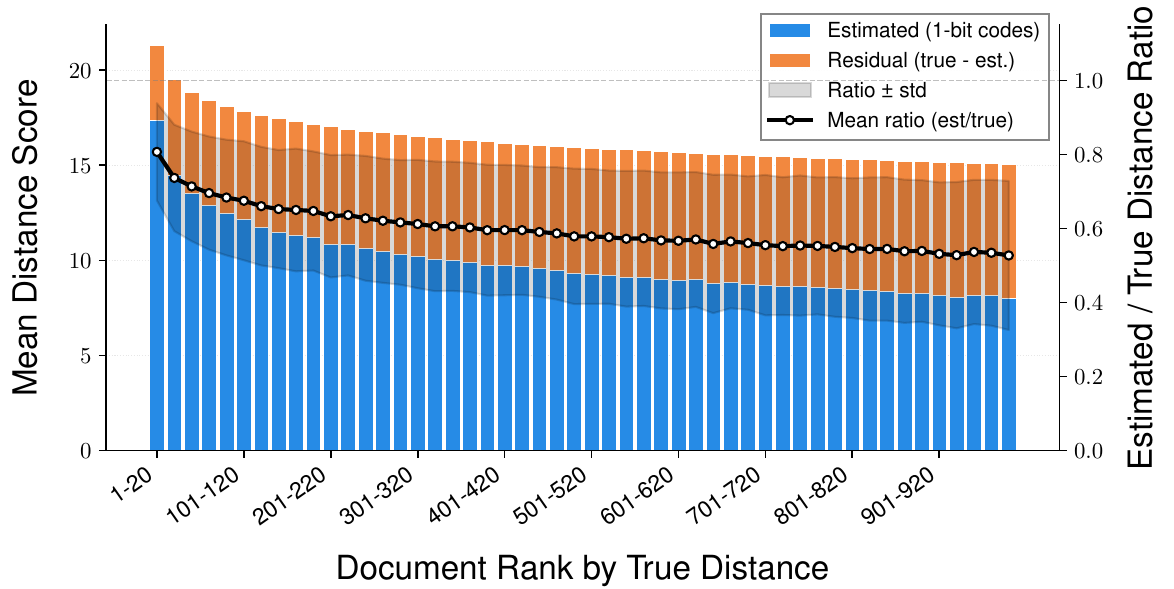}
        \caption{Candidate refinement: estimated vs true distance}
        \label{fig:refSub1}
    \end{subfigure}
    \hfill
    \begin{subfigure}{0.48\linewidth}
        \centering
        \includegraphics[width=\linewidth]{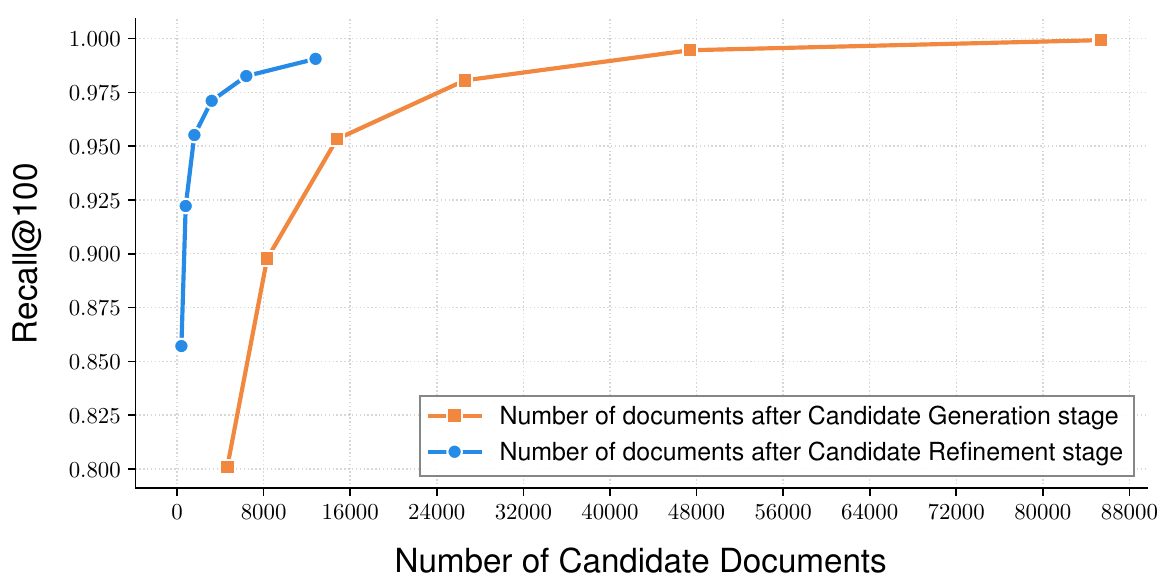}
        \caption{Recall vs number of documents, before and after refinement}
        \label{fig:refSub2}
    \end{subfigure}
    \vspace{-3mm}
    \caption{Effectiveness of the candidate refinement stage. 
    (a)~The ratio of estimated document distance (computed using 1-bit codes of retrieved tokens) to true distance is positively correlated with document ranking, indicating that the estimation effectively preserves the relative ordering of top candidates. (b)~The number of candidate documents required to achieve different Recall@100 levels before and after refinement. The candidate refinement stage aggressively prunes the search space, reducing the candidate set size by approximately $10\times$ while maintaining a $95\%$ recall target.}
    \label{exp:refinement}
\end{figure*}

\noindent
\textbf{\circled{2}~Candidate refinement} 
To achieve high recall, the candidate generation stage must retrieve a relatively large number of documents. During candidate refinement (CR) we reduce the size of this candidate set before the expensive document scoring stage.
The key observation is that document tokens do not contribute equally to the final document score. Tokens that are closest to the query tokens typically dominate the score. To validate this observation, we conduct the following experiment. We estimate the distance between candidate documents and the query using only the tokens retrieved during the candidate generation stage. Since this stage runs on GPU, the estimation uses only the $1$-bit quantization codes.

Figure~\ref{fig:refSub1} compares the estimated document distance computed from retrieved tokens with the true document distance computed using all tokens, and also plots their ratio. The results show a clear positive correlation between this ratio and the document ranking: higher-ranked documents tend to have larger ratios. Consequently, the estimated distance closely follows the trend of the true distance, making it effective for coarse ranking of candidate documents.

Figure~\ref{fig:refSub2}  shows the number of candidates required to achieve different recall levels. Even with $1$-bit quantization codes, CR reduces the candidate set size by roughly $10\times$ to achieve $95\%$ recall.

Formally, given the list of tokens $L_q$ retrieved by the IVF index in the CG stage for each query token $q\in Q$ and the candidate document set $S$ formed from the corresponding source documents, we estimate the (partial) score of a candidate document $V\in S$ as
\[
\mathcal{F'}(Q,V)=\sum_{q\in Q}\begin{cases}
                                    \max\limits_{v\in V\cap L_q}\left<q,v\right>, & V\cap L_q \neq \emptyset \\
                                    0\ , & V\cap L_q = \emptyset
                                \end{cases}
\]
In other words, we only compute contributions from tokens that were retrieved during candidate generation. 
Since tokens that are not retrieved contribute little to the final score, ignoring them significantly reduces the number of distance computations. 
Algorithm~\ref{alg:refinement} describes the refinement procedure in detail.

\begin{algorithm}[!t]
	\caption{Candidate Refinement} 
	\label{alg:refinement}
    {\small
        \KwIn{Query $Q$, list of token ids returned by the CG stage $L_q$ for each $q\in Q$, set of documents $S$ that tokens in $\{L_q\}_{q\in Q}$ belong to, target refinement size $k_{\mathsf{refine}}$}
        \KwOut{Refined candidate document set $S_\mathsf{refine}$}
        \BlankLine
        \For{$\mathsf{each}$ $\mathsf{doc}\in S$} {
            \For{$\mathsf{each}$ $q\in Q$} {
                Initialize $\mathsf{doc.score}[q]$ to be $0$
            }
        }
        \For{$\mathsf{each}$ $q\in Q$} {
            \For{$\mathsf{each}$ $i$ $\mathsf{from}$ $1$ $\mathsf{to}$ $|L_q|$} {
                $\mathsf{token\_id} \leftarrow L_q[i]$ \\
                $\mathsf{doc} \leftarrow \mathsf{get\_document}(\mathsf{token\_id})$ \\
                $\mathsf{dist} \leftarrow \mathsf{1\_bit\_distance}(\mathsf{token\_id}, q)$ \\
                $\mathsf{doc.score}[q]\leftarrow \max\{\mathsf{dist}, \mathsf{doc.score}[q]\}$
            }
        }
        \For{$\mathsf{each}$ $\mathsf{doc}\in S$} {
            $\mathsf{doc.score}\leftarrow$ sum of all $\mathsf{doc.score}[q]$ for each $q\in Q$
        }
        \Return $S_\mathsf{refine}\leftarrow$ top-$k_{\mathsf{refine}}$ documents with highest score
    }
\end{algorithm}

\begin{algorithm}[!t]
	\caption{Document scoring} 
	\label{alg:scoring}
    {\small
        \KwIn{Query $Q$, set ($S_{\text{refine}}$) of $k_\mathsf{refine}$ document ids returned by CR, number of chunks $c$, full-bit refinement size $k_\mathsf{full\text{-}bit}$}
        \KwOut{Top-$k$ documents}
        \BlankLine
        \For(\tcc*[f]{GPU processing}){$\mathsf{each}$ $i$ $\mathsf{from}$ $1$ $\mathsf{to}$ $c$} {
            $L_i\leftarrow$ the $i$'th chunk of $S_{\text{refine}}$ \\
            $D_i\leftarrow\mathsf{1\_bit\_document\_score}(L_i)$ \tcc*[f]{async GPU kernel} \\
            Asynchronously transfer $D_i$ from GPU to CPU            
        }
        $\mathsf{seen\_docs}\leftarrow\emptyset$ \\
        \For(\tcc*[f]{CPU processing}){$\mathsf{each}$ $i$ $\mathsf{from}$ $1$ $\mathsf{to}$ $c$} {
            $L_i\leftarrow$ the $i$'th chunk of $S_{\text{refine}}$ \\
            Wait on $D_i$ to arrive CPU \\
            $\mathsf{cand}\leftarrow$ top-$\frac{i}{c}k_\mathsf{full\text{-}bit}$ 1-bit score documents in $\bigcup\limits_{k=0}^i L_k$ \\
            $\mathsf{cand}\leftarrow \mathsf{cand}\backslash \mathsf{seen\_docs}$ \\
            compute full-bits document scores for $\mathsf{cand}$ \\
            $\mathsf{seen\_docs}\leftarrow \mathsf{seen\_docs}\cup \mathsf{cand}$ \\
        }
        \Return top-$k$ full-bits score documents in $\mathsf{seen\_docs}$
    }
\end{algorithm}

\stitle{\circled{3}~Document scoring} The document scoring stage takes the $k_\mathsf{refine}$ refined candidates, computes their full document scores, and returns the final top-$k$ results. To avoid expensive transfers of quantization codes from CPU to GPU, the full-bit distance computation is performed exclusively on the CPU.

However, since the CPU is significantly slower than the GPU, evaluating full-bit scores for all $k_\mathsf{refine}$ candidates is prohibitively expensive. Instead, \name first uses the GPU to compute low-precision scores based on 1-bit codes, and then uses these scores to filter the candidates. 
In contrast with the filtering that occurs during the candidate refinement stage, the GPU at this stage computes the complete 1-bit score (as opposed to the partial score $\mathcal{F}'$), using contributions from all tokens. The goal is to filter the candidate set more aggressively, now using more precise scores, before sending the reduced set to the CPU.
The CPU subsequently computes full-bit scores only for top-$k_\mathsf{full\text{-}bit}$ documents ranked by their $1$-bit scores. While this approach substantially reduces CPU workload, it introduces a strict dependency: the CPU must wait for the GPU to finish before it can begin, preventing any overlap between CPU and GPU computation and leading to underutilized resources.

To address this issue, we propose a CPU–GPU concurrent collaborative scoring strategy that achieves both limited CPU workload and effective computation overlap.
Algorithm~\ref{alg:scoring} shows the details of this approach. The algorithm splits the $k_\mathsf{refine}$ documents into $c$ chunks. GPU processes these chunks sequentially, while CPU processes earlier chunks concurrently with GPU processing later ones. Specifically, once the GPU finishes computing the $1$-bit scores for chunk $i$, it transfers the results to the CPU. After receiving them, the CPU ranks all documents seen so far based on their $1$-bit scores and computes the full-bit scores for documents that fall within the current top-$\frac{i}{c}k_\mathsf{full\text{-}bit}$ set but have not yet been processed. This design allows CPU computation for earlier chunks to overlap with GPU computation for later chunks, while still restricting the number of full-bit evaluations.

\begin{figure}[!t]
	\centering
	\includegraphics[width=0.4\textwidth]{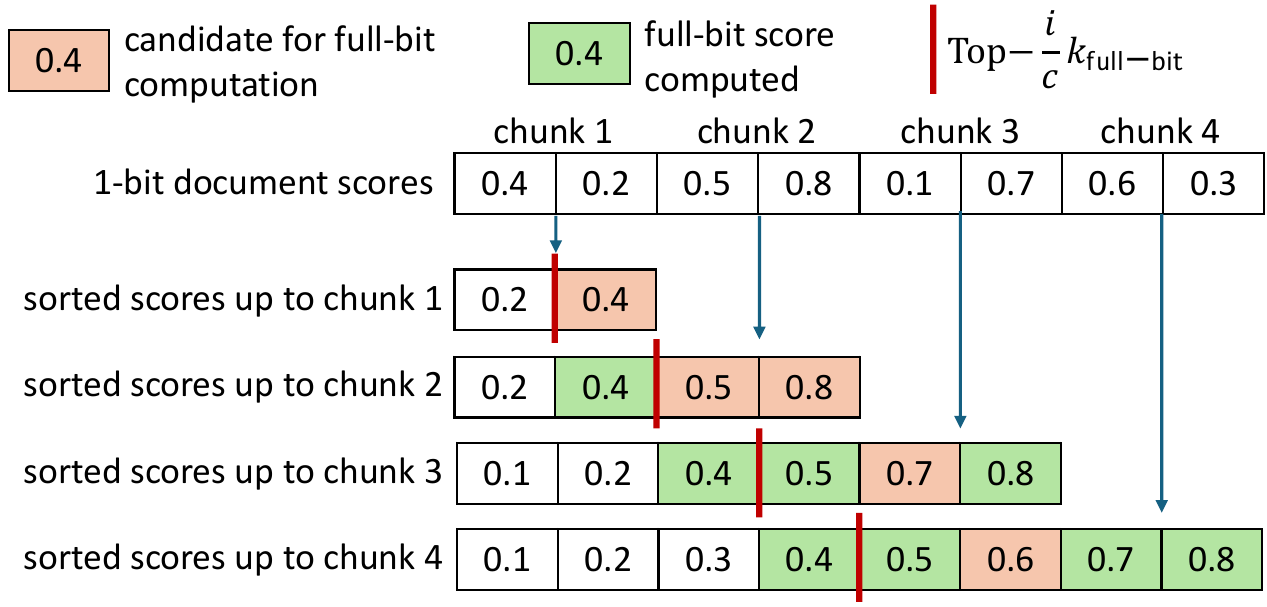}
	\caption{
    Illustration of CPU–GPU collaborative scoring. 
    Documents are divided into chunks and processed sequentially on the GPU to obtain $1$-bit scores (top row). After each chunk, the CPU maintains the sorted scores seen so far and computes full-bit scores for candidates that fall within the current top-$\frac{i}{c}k_{\text{full-bit}}$ threshold (indicated by red lines). Orange boxes denote candidates selected for full-bit computation, and green boxes denote documents whose full-bit scores have been computed.
    }
	\label{ex:cs}
	\Description{}
\end{figure}

Note that the algorithm always produces results at least as accurate as the first, non-concurrent scoring strategy, although it may compute slightly more full-bit scores. Figure~\ref{ex:cs} illustrates this behavior with an example. In this case, eight documents are evaluated using four chunks, and $k_{\text{full-bit}} = 4$, meaning that the top-4 documents according to the $1$-bit scores are refined using full-bit quantization codes. As shown, the algorithm correctly identifies and computes the full-bit scores for the true top-4 documents, while additionally evaluating one extra document beyond the strict top-$k_{\text{full-bit}}$ set. In the worst case, when documents appear in strictly increasing order of $1$-bit score across chunks, the total number of full-bit computations becomes
\[
\sum_{i=1}^{c} \frac{i}{c}k_\mathsf{full\text{-}bit}
=
\frac{c+1}{2}k_\mathsf{full\text{-}bit}.
\]
Such a skewed ordering is unlikely in practice, and the actual number of full-bit computations typically remains close to $k_\mathsf{full\text{-}bit}$.

\rthree{
\stitle{Parameter tuning}
We recommend to tune \name's retrieval parameters by working backward through the filtering pipeline. First, $n_{\mathrm{probe}}$ and $k_{\mathrm{refine}}$ should be set sufficiently large to prevent the candidate generation and refinement stages from prematurely discarding relevant documents. Next, $k_{\mathrm{full\text{-}bit}}$ should be increased until the desired recall is reached. Finally, $k_{\mathrm{refine}}$ and $n_{\mathrm{probe}}$ can be progressively reduced, in that order, to eliminate unnecessary computation and lower latency while preserving the target recall.
}

\subsection{Discussion and extensions}\label{sec:discussion}
\noindent\textbf{Scaling to larger datasets.}
\name assumes that the IVF-related index data and the $1$-bit quantization codes fit in a single GPU. For extremely large datasets, this assumption may no longer hold. In this case, \name can be naturally extended to a multi-GPU setting by partitioning the dataset across GPUs. Each GPU stores a disjoint shard of the IVF posting lists and corresponding $1$-bit codes, while the full-bit codes remain in CPU memory. At query time, query embeddings are broadcast to all GPUs, which independently perform candidate generation and refinement on their local shards. The local candidates and $1$-bit scores are then merged through a global top-$k$ reduction to obtain $S_{\mathsf{refine}}$. The final scoring stage follows the same GPU--CPU collaborative strategy as in the single-GPU design. Since communication mainly involves query embeddings, document ids, and scores rather than vector codes, this extension preserves \name's low data-movement advantage while scaling beyond a single GPU.

\rtwo{\stitle{Extending to other multi-vector settings}  \name focuses on token-level retrieval, where Chamfer similarity aggregates the maximum similarity of each query token over all document tokens.  However, its core design can also apply to settings where each entity has a set of distinct embeddings, such as text and image representations, and the score combines corresponding field-level similarities using a weighted sum or another fusion function~\cite{milvus,vbase,deg}. Each embedding field can maintain a lightweight GPU-resident index and low-precision codes for candidate generation, while higher-precision representations remain in CPU memory for final aggregate scoring. \name's GPU-based filtering can also be integrated with multi-index aggregation algorithms~\cite{NRA,vbase}, which combine the ranked results produced by the individual embedding indexes. Although token-specific techniques such as partial Chamfer scoring require adaptation, the general principle of GPU-based filtering followed by CPU-based high-precision ranking remains applicable.}

\stitle{Handling updates}
Practical applications may need to insert or delete multi-vectors from the dataset, which requires updating the IVF posting lists, CAGRA index, and the associated quantization-code storage in \name. Existing dynamic indexing techniques for IVF and graph indexes can be adopted to handle such updates~\cite{liu2025wolverine,singh2021freshdiskann,xu2023spfresh}. This is because \name uses IVF lookup as a subroutine, and the quantization codes are stored as per-token records associated with document ids. For insertions, new tokens can be quantized, assigned to their nearest IVF centroids, and appended to the corresponding posting lists, while their $1$-bit and full-bit codes are inserted into GPU and CPU code storage, respectively. For deletions, \name can maintain a deletion bitmap over document ids and skip deleted documents during retrieval, with periodic compaction or rebuilding when stale entries accumulate. Therefore, \name can support practical update requirements without changing its query processing pipeline.

%% file: sections/impl-details.tex
\section{Kernel optimizations in \name}\label{sec:sys-opt}

In this section, we present \name's hardware-aware optimizations, designed to maximize GPU efficiency. \rtwo{The main computation kernels in \name are highly memory-bound: they perform relatively simple arithmetic, such as binary inner products, maximum updates, and reductions, over large token lists and score arrays. Consequently, their performance is primarily limited not by arithmetic throughput, but by how efficiently data can be moved through the GPU memory hierarchy. The primary goal of our kernel optimizations is therefore to improve memory access efficiency---by increasing data reuse, improving memory coalescing, reducing unnecessary global-memory traffic, and mitigating synchronization overhead.} First, we introduce a LUT-based optimization for binary inner product computation that eliminates branch divergence and improves shared-memory efficiency through query tiling (Section~\ref{sec:lut}). Second, we describe specialized kernels for document score aggregation, which adopt stage-specific designs---fused atomic updates for candidate refinement and decoupled shared-memory reductions for document scoring---to mitigate atomic contention and memory bottlenecks (Section~\ref{sec:doc-score}).

\subsection{LUT-based binary inner product}\label{sec:lut}
One core operation in our retrieval pipeline is computing the inner product between a floating-point query token and binary-encoded document tokens. A direct implementation scans the code bit by bit and conditionally accumulates the corresponding query dimensions, which introduces data-dependent branching in the inner loop and is therefore inefficient on GPUs. \rtwo{This inefficiency is particularly harmful because GPUs execute threads in groups, called warps, where threads are most efficient when they follow the same instruction path. When different bits trigger different conditional branches across threads, the warp must serialize these paths, reducing parallel efficiency.} We instead adopt a widely-used LUT-based optimization~\cite{fastscan}. Let $b=(b_0,\dots,b_{d-1})$ denote a 1-bit code, and partition it into $m=d/4$ chunks of four consecutive bits $b^{(j)}\in\{0,1\}^4$. For each query token $q$, we precompute
$T_q[j,u] = \sum_{t=0}^{3} u_t \, q_{4j+t}$, $u \in \{0,1\}^4$,
so that the binary inner product becomes
$\langle q,b\rangle = \sum_{j=0}^{m-1} T_q[j,b^{(j)}]$.
This replaces $d$ bit-wise accumulations with $d/4$ table lookups and additions while also eliminating branch divergence.

\begin{figure}[t]
    \centering
    \begin{subfigure}{0.48\linewidth}
        \centering
        \includegraphics[width=0.85\linewidth]{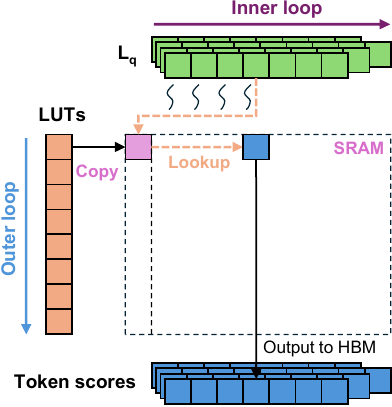}
        \caption{Candidate refinement stage}
        \label{fig:ipSub1}
    \end{subfigure}
    \hfill
    \begin{subfigure}{0.48\linewidth}
        \centering
        \includegraphics[width=0.85\linewidth]{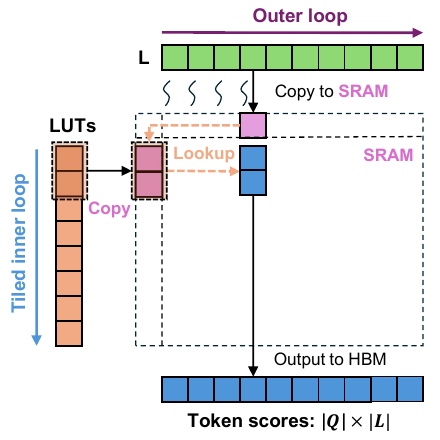}
        \caption{Document scoring stage}
        \label{fig:ipSub2}
    \end{subfigure}
    \vspace{-3mm}
    \caption{\looseness-1
    Illustration of the LUT-based binary inner product kernels on the GPU. 
    (a)~During candidate refinement, a thread block loads a single query token's LUT into shared memory (SRAM) and computes inner products against its specific candidate list of document tokens ($L_q$). (b)~During document scoring, document tokens ($L$) are loaded into SRAM and scored against all $|Q|$ query tokens. To alleviate SRAM pressure, the computation tiles the query LUTs in the inner loop.
    }
    \label{fig:ip-kernels}
\end{figure}

We realize the LUT-based approach on GPU as follows. During candidate refinement, each query token probes its own set of nearby clusters, so the list of document tokens it needs to compare against is specific to that query token. We therefore dedicate one thread block to the inner product computation between one query token and a group of document tokens (illustrated in Figure~\ref{fig:ipSub1}): the block first loads that query's LUT into shared memory (SRAM), then each thread computes one document token's inner product against the query token using the LUT. \rtwo{Shared memory is a small but low-latency on-chip memory managed explicitly by the kernel. Placing the LUT in shared memory avoids repeatedly fetching the same query-dependent table from the much slower global memory, allowing many threads in the block to reuse it efficiently.} One complication is that the binary codes are stored in cluster-oriented layout (i.e., token codes in the same cluster are placed together), which means that tokens belonging to the same document are scattered across memory, resulting in uncoalesced access. \rtwo{On GPUs, memory bandwidth is highest when neighboring threads in a warp access neighboring memory addresses, allowing their requests to be coalesced into a small number of memory transactions. In contrast, scattered accesses generate many separate transactions and underutilize memory bandwidth.} To improve coalescing, we optionally allow pre-storing an additional copy of the binary codes reordered by document assignment, ensuring that document tokens within the same cluster are contiguous in memory. This layout allows a warp scanning a document to issue coalesced $d$-bit code loads. 

\looseness-1
In the document scoring stage, every surviving document token must be scored against all $|Q|$ query tokens, and the set of document tokens is shared across queries. We therefore assign each thread to compute the inner product of one token against all query tokens, as illustrated in Figure~\ref{fig:ipSub2}. This allows us to load one token's binary code once and reuse across the computation of all query tokens. It is also desirable for each thread block to load LUTs of all query tokens into SRAM. However, there will be a total of $|Q|\times m\times 16$ LUT entries, which takes 64KB when $|Q|=32$ and $d=128$. The large LUT size imposes high pressure on SRAM and thus results in resource contention. \rtwo{This pressure is not only a capacity issue: excessive shared-memory usage can also reduce kernel occupancy, i.e., the number of thread blocks that can reside concurrently on a streaming multiprocessor, thereby limiting the GPU's ability to hide memory latency.} To address this issue, we instead tile along the query dimension: the $Q$ queries are split into $Q/\mathtt{TILE\_Q}$ tiles, and for each tile the block loads only that tile's slice of the LUT into shared memory. Every thread in the block then fetches its document token's 4-bit chunks a single time into registers and reuses them across all $\mathtt{TILE\_Q}$ queries of the tile before advancing to the next tile. \rtwo{This tiling strategy preserves the reuse benefits of shared memory while keeping the per-block working set small enough to maintain high parallelism.}

\subsection{Document score aggregation}\label{sec:doc-score}
After computing the token-level inner products, we aggregate them into a single score per document using the Chamfer rule. Specifically, for each query token, we take the maximum inner product over all tokens in the document (the aggregation step), and then sum these per-query maxima (the summation step). While the summation step can be straightforwardly performed using a parallel tree reduction kernel, the candidate refinement and document scoring stages differ in how the aggregation step is carried out, as they operate on working sets with distinct characteristics. 

In the document scoring stage, the aggregation pattern is regular. The goal is to compute document scores for a candidate list $L$, given a token score matrix of shape $[|Q|, N]$ produced by the inner product kernel, where $N=\sum_{V\in L}|V|$ is the total number of tokens across all candidate documents, and rows correspond to query tokens. We first assign each thread block to process a group of documents, with each thread responsible for one document. Concretely, each thread iterates over all query tokens; for each query token, it scans all tokens of the document to compute the maximum score. After obtaining the per-query maxima for all $|Q|$ query tokens, these values are summed using a parallel binary tree reduction to obtain the final document score. However, this direct approach suffers from severe memory bottleneck. Because consecutive threads are assigned to different documents within the group, their global memory accesses are scattered and poorly coalesced, leading to low memory bandwidth utilization. To address this, we decouple the memory loading phase from the maxima update phase. Specifically, threads within a block first cooperatively load the required token scores into shared memory using fully coalesced accesses, and then perform the per-query maxima updates. In addition, we adopt the same tiling strategy as in the binary kernel for the document scoring stage, which helps alleviate shared memory pressure and improves overall resource utilization.

\begin{algorithm}[!t]
    \caption{Atomic aggregation for CR stage}
    \label{alg:naive-cr-agg}
    {\small
        \KwIn{Retrieved token lists $\{L_q\}_{q\in Q}$, token scores $\{F_q\}_{q\in Q}$}
        \KwOut{Score table $M[d,q]$, initialized to $-\infty$}
        \For{$\mathsf{each}$ $q\in Q$ $\mathsf{in}$ $\mathsf{parallel}$}{
            \For{$\mathsf{each}$ $i\in\{1,\dots,|L_q|\}$ $\mathsf{in}$ $\mathsf{parallel}$}{
                $t \leftarrow L_q[i],\quad d \leftarrow \mathsf{get\_document}(t)$ \\
                $s \leftarrow F_q[i]$ \\
                $\mathsf{atomicMax}(M[d,q],\, s)$
            }
        }
    }
\end{algorithm}

\looseness-1
Candidate refinement, by contrast, exhibits a completely different aggregation pattern. Its input is the collection of retrieved token lists $\{L_q\}_{q\in Q}$ produced by the candidate generation stage. Tokens are organized by their cluster assignment rather than by source document, meaning that continuous tokens may belong to arbitrarily different documents. Therefore, a thread block that cooperatively processes a contiguous chunk of $L_q$ cannot safely perform a scan-based reduction to obtain per-document maxima. Updating the per-document-per-query score table $M[d,q]$ under such scattered writes fundamentally requires an atomic maximum operator.

Algorithm~\ref{alg:naive-cr-agg} depicts how the aggregation is performed in the candidate refinement stage: it launches one thread per entry of $\{L_q\}_{q\in Q}$, with each thread fetching the corresponding token score, and then issues an \texttt{atomicMax} against the corresponding slot of $M$.

While simple and highly parallel, this kernel may suffer from severe atomic contention. A document whose tokens lie in many of the clusters probed by $q$ appears multiple times in $L_q$, causing many threads to update the same memory slot $M[d,q]$. These concurrent \texttt{atomicMax} operations block each other, stalling the issuing warps and leaving execution units idle.

To alleviate this contention, we fuse the aggregation step directly into the binary inner product kernel rather than executing it as a separate pass. Under this design, each thread interleaves atomic updates with the inner-product computation of subsequent tokens, introducing tens of cycles of independent work between atomics. This reduces the temporal clustering of atomic requests compared to a dedicated aggregation kernel that issues them back-to-back, improving latency hiding and alleviating contention. Fusion also eliminates the need to materialize and reload the intermediate $\sum_q |L_q|$ token score array, reducing global memory traffic.

%% file: sections/experiments.tex
\section{Experimental evaluation}\label{sec:expr}
In this section, we present extensive experiments of our system and its components, and we contrast with the state of the art.  Our evaluation aims to answer the following research questions:

\begin{itemize}[leftmargin=10pt, topsep=0pt]

    \item \textbf{End-to-end performance:} How does \name compare to state-of-the-art multi-vector retrieval systems in terms of latency, throughput, and accuracy (Section~\ref{sec:end2end})?

    \item \textbf{Chunking strategy:} How does the overlapping chunk size affect system performance? How much redundant computation does overlapping introduce, more specifically, how does the actual number of full-bit computations compare to $k_{\text{fullbit}}$ (Section~\ref{sec:ablation})?
    
    \item \textbf{Optimizations:} What is the impact of kernel and algorithmic optimizations on end-to-end system performance (Section~\ref{sec:ablation})?

\end{itemize}

\subsection{Experimental settings}
\noindent\textbf{Implementation.} We implement \name in CUDA C++ with customized GPU kernels. We leverage the RaBitQ library~\cite{rabitq_library} for vector quantization and NVIDIA's cuVS library~\cite{rapidsai_cuvs} for CAGRA index construction and search. The code is compiled using C++17 and CUDA 12.9 compilers, with \texttt{-O3} and \texttt{-march=native} flags enabled.

\stitle{Datasets} We evaluate \name on three publicly available real-world, large-scale datasets. LoTTE Pooled~\cite{colbertv2} is a long-form text retrieval benchmark that aggregates diverse domains to evaluate generalization in multi-vector retrieval settings; HotpotQA~\cite{hotpot} is a multi-hop question-answering dataset requiring reasoning over multiple documents; and MS MARCO~\cite{msmarco} is a widely used large-scale passage ranking benchmark derived from real-world search queries, emphasizing relevance matching at scale. We use the ColBERTv2 embedding model~\cite{colbertv2} to generate the document and query multi-vectors with 128 dimensions.
The dataset statistics are summarized in Table~\ref{tab:datasets-and-indexes}.

\stitle{Baselines}
We compare \name against state-of-the-art multi-vector retrieval systems on CPU and GPU platforms. 
\textbf{IGP}~\cite{igp} \rmeta{and \textbf{GEM}~\cite{gem}} are two CPU-based approaches with highly-optimized graph-based search algorithms, serving as strong non-GPU baselines.
\textbf{PLAID}~\cite{plaid} is a state-of-the-art system with GPU support, making it the primary baseline for GPU-accelerated multi-vector retrieval. 
We further implement \textbf{PLAID+}, a variant of PLAID that keeps all index data (including quantized vectors) GPU-resident to include its performance without host--device transfer overhead.

\stitle{Platforms} Unless otherwise stated, experiments are conducted on a single server equipped with an NVIDIA Tesla V100-SXM2-32GB GPU, an Intel Xeon Gold 6240 CPU @ 2.60\,GHz, and 386\,GB of RAM, running Ubuntu 24.04.3 LTS, and use 32 CPU threads with hyper-threading disabled.

\begin{table}[!t]
    \centering
    {\small
    \begin{tabular}{lcccc}
        \toprule
        \textbf{Dataset} & \textbf{Queries} & \textbf{Documents} & \textbf{Tokens} & \textbf{Size} \\
        \midrule
        LoTTE Pooled & 2,931 & 2.4M & 266M & 136 GB \\
        HotpotQA & 1,000 & 5.2M & 283M & 146 GB \\
        MS MARCO & 6,980 & 8.8M & 597M & 308 GB \\
        \bottomrule
    \end{tabular}}
    \caption{Summary statistics of the experiment datasets }
    \label{tab:datasets-and-indexes}
    \vspace{-4mm}
\end{table}

\begin{figure*}[!t] 
    \centering
    \includegraphics[width=0.9\linewidth]{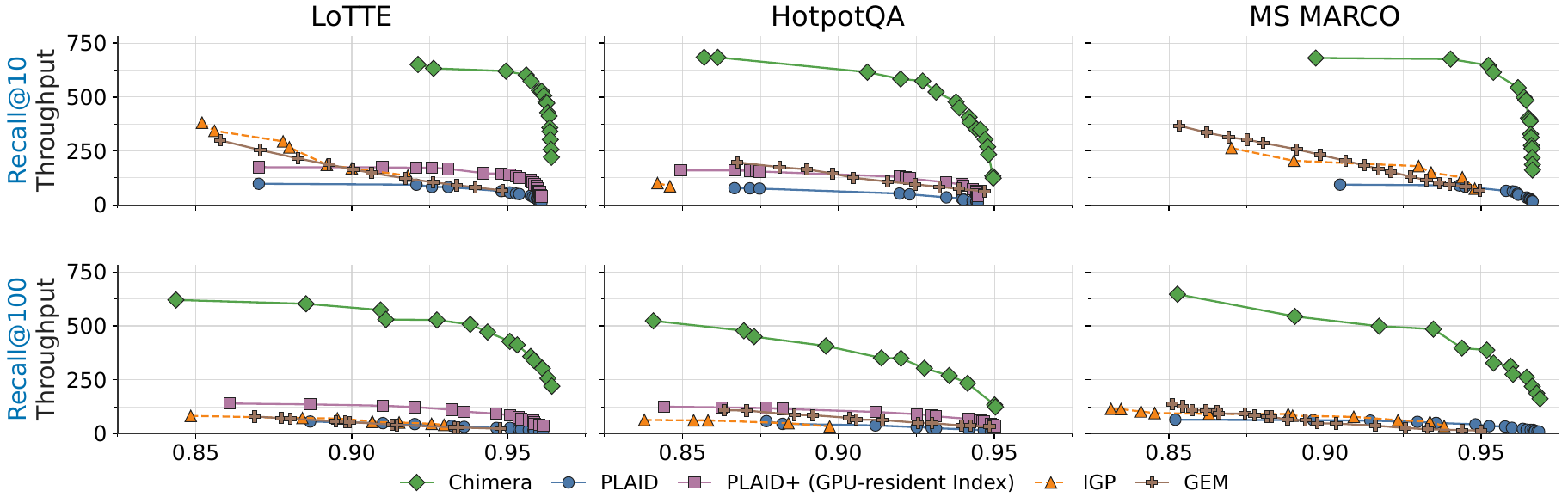}
    \vspace{-3mm}
    \caption{End-to-end performance comparison of \name with PLAID, PLAID+, IGP and GEM. PLAID+ is omitted on MS MARCO because its GPU-resident index exceeds available GPU memory.}
    \vspace{-3mm}
    \label{fig:end_to_end}
\end{figure*}

\stitle{Evaluation protocol} We are interested in both the efficiency and effectiveness of multi-vector retrieval. For efficiency, we use throughput (measured as queries per second) as the main performance metric. For effectiveness, we use Recall@$k$, which measures the fraction of ground-truth top-$k$ multi-vectors that are successfully retrieved within the returned top-$k$ results. \rone{Following PLAID’s evaluation protocol, we evaluate PLAID and \name one query at a time, whereas the CPU-based competitors process queries concurrently, assigning one query to each thread.} To ensure a fair comparison, we control all compared methods to use the same level of vector compression (4/8 bits per dimension). \rthree{Index parameters of baselines follow their original papers or source codes, and all competitors are tuned to their best query performance.}
We repeat each experiment 3 times and report the average as the final results.

\subsection{End-to-end performance comparison}
\label{sec:end2end}


Figure~\ref{fig:end_to_end} shows the end-to-end throughput and recall tradeoff of \name compared to PLAID, PLAID+, IGP and GEM. Across all datasets, \name consistently achieves higher throughput at matched recall. The advantage becomes more pronounced in the high-recall regime, where baseline systems degrade sharply.

Table~\ref{tab:speedup} summarizes the throughput\rone{/latency}\rone{\footnote{\rone{We evaluate PLAID and \name in a single-query, non-batched manner, so the reported QPS directly corresponds to mean per-query latency (latency $=1/\mathrm{QPS}$).}}} comparison with PLAID at fixed recall targets (0.90, 0.95, \rone{and 0.99}). \rone{Unless otherwise noted, our experiments use 4-bits-per-dimension quantization, which provides a favorable efficiency--memory tradeoff but limits the maximum attainable recall to around 0.95. To study more stringent quality targets, we additionally evaluate both systems with 8-bits-per-dimension quantization, which enables higher recall at the cost of larger index size and higher runtime memory consumption. These higher-recall settings are highlighted in gray in Table~\ref{tab:speedup}.}
Across all datasets and top-$k$ settings, \name delivers substantial improvements over PLAID, with speedups ranging from $6.95\times$ to \rone{$59.48\times$}. 
\rtwo{Notably, \name's gains over PLAID increase at higher recall settings, and PLAID struggles to achieve Recall@100$=0.99$, incurring substantially higher query processing time. This widening advantage stems from two factors. First, PLAID uses coarse IVF clusters with large, low-selectivity posting lists, requiring it to probe more clusters and scan more documents to avoid missing relevant results. In contrast, \name's fine-grained clustering carries fewer irrelevant documents into subsequent stages. Second, PLAID's scalar quantization is less accurate than \name's RaBitQ. As the nearest vectors become harder to distinguish at higher recall, PLAID must scan additional documents to compensate for quantization error. Together, these factors cause PLAID's workload to grow more sharply as the recall target increases, leading to a larger relative speedup for \name in stringent recall regimes.}

\begin{table}[t]
\centering
\resizebox{0.9\linewidth}{!}{
\begin{tabular}{lcccc}
\toprule
\textbf{Dataset (Top-$k$)} & \textbf{Target} & \textbf{PLAID QPS} & \textbf{\name QPS} & \textbf{Speedup} \\
\midrule
LoTTE ($k$=10)   & 0.90 & 93.6  & 650.7 & \textbf{6.95$\times$} \\
LoTTE ($k$=10)   & 0.95 & 57.8  & 602.4 & \textbf{10.42$\times$} \\
LoTTE ($k$=100)  & 0.90 & 51.5  & 573.7 & \textbf{11.13$\times$} \\
LoTTE ($k$=100)  & 0.95 & 26.8  & 428.2 & \textbf{15.97$\times$} \\
\rowcolor{lightgray}
LoTTE ($k$=100)  & 0.99   & 3.7 & 220.1 & \textbf{59.48$\times$} \\
\midrule
HotpotQA ($k$=10)  & 0.90 & 52.6  & 615.5 & \textbf{11.70$\times$} \\
HotpotQA ($k$=10)  & 0.95 & 9.1$^\dagger$  & 123.9 & \textbf{13.62$\times$} \\
HotpotQA ($k$=100) & 0.90 & 38.3  & 351.4 & \textbf{9.17$\times$} \\
HotpotQA ($k$=100) & 0.95 & 9.3   & 133.6 & \textbf{14.32$\times$} \\
\rowcolor{lightgray}
HotpotQA ($k$=100) & 0.99   & 3.7 & 134.4 & \textbf{36.32$\times$} \\
\midrule
MS MARCO ($k$=10)  & 0.90 & 93.9  & 676.0 & \textbf{7.20$\times$} \\
MS MARCO ($k$=10)  & 0.95 & 65.0  & 646.7 & \textbf{9.95$\times$} \\
MS MARCO ($k$=100) & 0.90 & 60.6  & 498.5 & \textbf{8.23$\times$} \\
MS MARCO ($k$=100) & 0.95 & 35.2  & 387.7 & \textbf{11.01$\times$} \\
\rowcolor{lightgray}
MS MARCO ($k$=100) & 0.99   & 8.5 & 216.6 & \textbf{25.48$\times$} \\
\bottomrule
\end{tabular}
}
\caption{
Throughput comparison for PLAID and \name.
The $^\dagger$ reports QPS at the recall closest to 0.95 achieved by PLAID (0.9447) on HotpotQA for $k=10$.
\rone{The gray-highlighted rows display results at 8-bit-per-dimension quantization, which allows us to evaluate the systems at 0.99 recall level.}
}
\vspace{-2mm}
\label{tab:speedup}
\end{table}


%
These improvements stem from \name's ability to reduce the cost of the dominant stages in the retrieval pipeline. As shown in Section~\ref{sec:ablation} (Figure~\ref{fig:lotte_latency_breakdown}), end-to-end latency is primarily dominated by candidate refinement and document scoring. \name accelerates these stages through optimized GPU kernels and a chunked execution strategy that overlaps GPU-side candidate refinement with CPU-side document scoring, thereby reducing exposed latency and improving throughput. 


\begin{table}[t]
\centering
{\rthree{\small
\begin{tabular}{@{\extracolsep{\fill}}lcccccc@{}}
\toprule
& & & \multicolumn{4}{c}{\textbf{latency (ms)}} \\
\cmidrule(lr){4-7}
\textbf{Dataset} & \textbf{System} & \textbf{Recall} & \textbf{Mean} & \textbf{p50} & \textbf{p95} & \textbf{p99} \\
\midrule
\multirow{4}{*}{\textbf{LoTTE}}
 & \name  & 0.9  & \textbf{1.9} & \textbf{1.9} & \textbf{2.0} & \textbf{2.1} \\
 & PLAID  & 0.9  & 12.6 {\footnotesize(6.7$\times$)} & 11.9 & 17.3 & 23.8 \\
 & \name  & 0.95 & \textbf{2.4} & \textbf{2.4} & \textbf{2.5} & \textbf{2.6} \\
 & PLAID  & 0.95 & 29.7 {\footnotesize(12.5$\times$)} & 28.7 & 39.2 & 50.2 \\
\midrule
\multirow{4}{*}{\textbf{HotpotQA}}
 & \name  & 0.9  & \textbf{2.5} & \textbf{2.5} & \textbf{2.7} & \textbf{2.9} \\
 & PLAID  & 0.9  & 19.3 {\footnotesize(7.7$\times$)} & 16.7 & 26.9 & 37.7 \\
 & \name  & 0.95 & \textbf{5.5} & \textbf{5.5} & \textbf{6.2} & \textbf{6.5} \\
 & PLAID  & 0.95 & 81.0 {\footnotesize(14.6$\times$)} & 79.8 & 100.0 & 115.5 \\
\midrule
\multirow{4}{*}{\textbf{MS MARCO}}
 & \name  & 0.9  & \textbf{1.8} & \textbf{1.8} & \textbf{1.9} & \textbf{2.0} \\
 & PLAID  & 0.9  & 13.0 {\footnotesize(7.3$\times$)} & 12.1 & 18.4 & 25.9 \\
 & \name  & 0.95 & \textbf{2.2} & \textbf{2.2} & \textbf{2.4} & \textbf{2.5} \\
 & PLAID  & 0.95 & 19.8 {\footnotesize(9.0$\times$)} & 18.2 & 28.3 & 39.7 \\
\bottomrule
\end{tabular}}
}
\caption{\rthree{Single-query latency distributions for \name and PLAID on A100.}}
\vspace{-2mm}
\label{tab:a100_latency}
\Description{}
\end{table}

\rthree{To demonstrate that the benefits of \name's design generalize across GPU platforms, we further report single-query latency distributions on an A100 GPU in Table~\ref{tab:a100_latency}. \name consistently achieves lower mean, p50, p95, and p99 latency than PLAID, reducing mean latency by $6.7\times$--$14.6\times$ across datasets and recall targets. This shows that the advantage persists on a modern GPU platform with a faster CPU--GPU interconnect than V100.}

\rone{Finally, Table~\ref{tab:mrr_ndcg} reports MRR@10 and NDCG@10 for \name and PLAID on the three datasets. \name consistently outperforms PLAID at the same quality level across all datasets, demonstrating its effectiveness in retrieval-oriented ranking tasks.}

Overall, \name improves not only individual components but the full end-to-end retrieval pipeline, enabling substantially higher throughput at a fixed quality target.

\subsection{Analysis of \name's performance}
\label{sec:ablation}
\begin{figure}[!t]
  \centering
  \includegraphics[width=\linewidth]{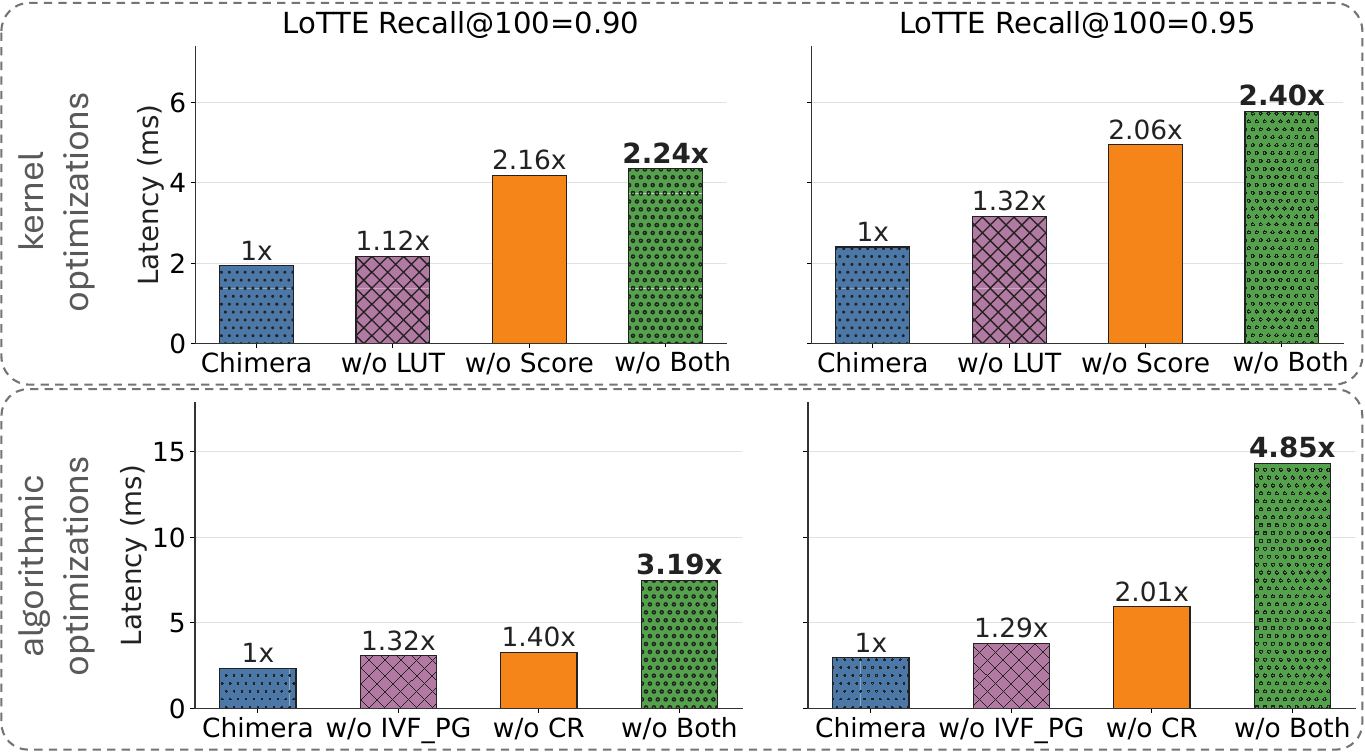}
  \vspace{-6mm}
  \caption{
  \rthree{Impact of \name's optimizations on query latency on LoTTE, evaluated at target Recall@100 levels of 0.90 and 0.95. The top graphs demonstrate the impact of the kernel optimizations (LUT and document score aggregation), and the bottom row evaluate the algorithmic designs (combined IVF and proximity graph, and candidate refinement.}
  }
  \label{fig:v8_variant_targets_latency}
\end{figure}

\noindent\textbf{Impact of kernel optimizations.}
\looseness-1
We evaluate the impact of \name's two kernel-level optimizations---lookup table (LUT) optimization and document score aggregation---through ablation. The top graphs of Figure~\ref{fig:v8_variant_targets_latency} report the average query latency on LoTTE for the full system and three variants: one without LUT optimization, one without document score aggregation, and one with both disabled. For each variant, we report its slowdown relative to the full system.
Both optimizations reduce query latency, although document score aggregation has the larger impact. Disabling it increases latency by $2.06\times$--$2.16\times$, demonstrating that efficiently accumulating token-level scores into document-level scores is critical for this workload. Disabling LUT optimization results in a smaller but still noticeable slowdown of $1.12\times$--$1.32\times$, confirming the benefit of accelerating the underlying distance computations. When both optimizations are disabled, latency increases by $2.24\times$--$2.40\times$. These results show that the two optimizations are complementary, with document score aggregation providing the dominant benefit and LUT optimization further reducing latency across both recall targets.

\begin{table}[t]
\centering
\rone{\small
\begin{tabular}{@{}lcccc@{}}
\toprule
\textbf{Dataset} & \textbf{System} & \textbf{MRR@10} & \textbf{NDCG@10} & \textbf{QPS} \\
\midrule
\multirow{2}{*}{\textbf{LoTTE}}   & \name   & 0.362 & 0.412     & 408.2      \\
                              & PLAID   & 0.361 & 0.411     & 46.3      \\
\hline
\multirow{2}{*}{\textbf{HotpotQA}}& \name   & 0.708 & 0.743     & 331.2      \\
                              & PLAID   & 0.706 & 0.741 & 25.4      \\
\hline
\multirow{2}{*}{\textbf{MS MARCO}}& \name   & 0.388 & 0.459 & 489.9      \\
                              & PLAID   & 0.388 & 0.460     & 58.3      \\
\bottomrule
\end{tabular}}
\caption{\rone{MRR and NDCG of \name and PLAID on LoTTE, HotpotQA and MS MARCO datasets.}}
\vspace{-2mm}
\label{tab:mrr_ndcg}
\Description{}
\end{table}

\rthree{
\stitle{Impact of algorithmic designs}
We further evaluate the two main algorithmic components of \name: the use of fine-grained IVF clusters together with a proximity graph over the cluster centroids (IVF\_PG), and the candidate refinement (CR) stage. The bottom graphs of Figure~\ref{fig:v8_variant_targets_latency} display the results. For the non-IVF\_PG variants, we use the same cluster count adopted by PLAID. At Recall@100$=0.90$, removing IVF-PG and candidate refinement increases latency by $1.32\times$ and $1.40\times$, respectively. Their benefits become more pronounced at Recall@100$=0.95$: disabling candidate refinement incurs a $2.01\times$ slowdown, while disabling IVF-PG incurs a $1.29\times$ slowdown. Candidate refinement is particularly important at higher recall because the larger candidate set produced under more stringent quality requirements makes it increasingly costly to send all candidates directly to complete document scoring.

The two designs are also complementary. Disabling both increases latency by $3.19\times$ and $4.85\times$ at Recall@100$=0.90$ and $0.95$, respectively---substantially more than disabling either component alone. Fine-grained clustering reduces the number of irrelevant tokens and documents introduced during candidate generation, while candidate refinement further prunes the surviving documents using inexpensive partial scores before complete document scoring. When either design is present, it can partially compensate for the absence of the other; when both are removed, substantially more documents reach the expensive final stage. These results demonstrate that \name's performance gains do not arise solely from its heterogeneous memory placement or avoidance of CPU--GPU transfer. Its algorithmic filtering pipeline substantially reduces the amount of scoring computation, helping explain why \name can outperform a fully GPU-resident baseline such as PLAID+ even when the dataset fits in GPU memory.
}

\begin{figure}[t]
  \centering
  \includegraphics[width=\linewidth]{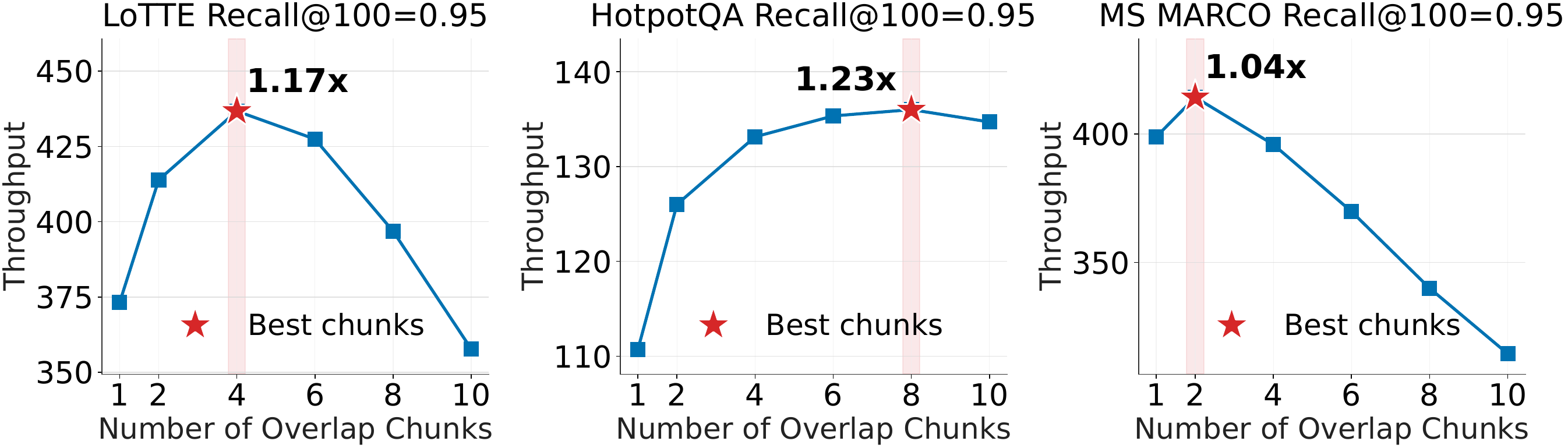}
  \vspace{-6mm}
  \caption{Impact of overlap chunk count on throughput. 
  }
  \label{fig:qps_vs_overlap_topk100_recall0.95}
\end{figure}


\stitle{Impact of chunking strategy}
In this experiment, we study the impact of the number of chunks~($c$) in the GPU--CPU collaborative scoring strategy. Figure~\ref{fig:qps_vs_overlap_topk100_recall0.95} reports the throughput of \name under different values of $c$ on LoTTE, HotpotQA, and MS MARCO at target Recall@100 levels of $0.95$. \rone{These results isolate the incremental benefit of chunking within collaborative scoring; the broader architectural benefit of avoiding document-vector transfer is evaluated separately below.} Overall, chunking consistently improves throughput over the non-overlapped setting ($c=1$), confirming the benefit of overlapping CPU full-bit scoring with GPU 1-bit scoring, although too many chunks can hurt performance due to increased scheduling and synchronization overhead and reduced GPU efficiency. The optimal chunk count depends on the balance between CPU and GPU workloads: when GPU work dominates, overlap yields limited benefit (e.g., MS MARCO, where $c=2$ is best with only a marginal gain of $1.04\times$ over $c=1$), whereas when both CPU and GPU workloads are substantial, more chunks better overlap execution (e.g., on HotpotQA, where $c=8$ performs best, achieving speedup of $1.23\times$). For LoTTE, moderate chunk counts ($c=4$) achieve the best performance, improving throughput by $1.17\times$ over the single-chunk baseline. \rone{The end-to-end benefit of chunking also depends on the fraction of total query-processing time occupied by document scoring. The modest gain on MS MARCO arises because relatively few documents reach the document-scoring stage, whereas the benefit is substantially larger when this stage carries a heavier workload: on LoTTE at Recall@100$=0.99$, chunking improves throughput by $42\%$.}

We further evaluate whether chunking introduces additional CPU computation by measuring the actual number of full-bit document scores, denoted by $\hat{k}_\mathsf{full\text{-}bit}$, with $k_\mathsf{full\text{-}bit}=400$ and $k_\mathsf{refine}\in\{2000,4000,8000\}$. On LoTTE, both the average and maximum remain exactly $400$ in all settings. On HotpotQA, even at $k_\mathsf{refine}=8000$, only three out of one thousand queries exceed $400$ computations; the average is $400.098$ and the maximum is $441$. On MS MARCO, the average never exceeds $400.011$ and the maximum is at most $440$. Thus, although chunking can increase the worst-case computation, such cases are rare and the overlap introduces almost no additional full-bit scoring in practice.

\stitle{Data transfer}
One of the key advantages of \name is that it eliminates the need for expensive data transfer between CPU and GPU. Table~\ref{tab:transfer} verifies it by comparing the latency, transfer time, and transfer size of \name with PLAID across three datasets at Recall@100=0.9. PLAID incurs substantial transfer overhead because it must upload vector data from CPU memory to the GPU for scoring at query time, moving 28.1--56.1 MiB of data per query and spending 6.6--7.7 ms on data transfer, which accounts for a large fraction of its end-to-end latency. In contrast, thanks to its design of keeping compact $1$-bit quantization codes and index data resident on the GPU, \name only needs to transfer hundreds of document ids and scores in the document scoring stage, reducing the transfer size to only 0.04--0.09 MiB per query. As a result, its transfer time is nearly negligible, remaining below 0.08 ms across all datasets. These results demonstrate that \name effectively removes CPU--GPU data movement as a performance bottleneck.

\begin{table}[t]
\centering
\resizebox{0.9\columnwidth}{!}{
\begin{tabular}{@{}lcccc}
\toprule
\textbf{Dataset} & \textbf{Method} & \makecell[tc]{\textbf{Latency}\\\textbf{(ms)}} & \makecell[tc]{\textbf{Transfer time}\\\textbf{(ms)}} & \makecell[tc]{\textbf{Transfer size}\\\textbf{(MiB)}} \\
\midrule
\multirow{2}{*}{\textbf{LoTTE}}   & PLAID & 19.4 & 6.6\phantom{1} & 30.4\phantom{1}           \\
                                  & \name & \phantom{1}1.9 & 0.04 & 0.04           \\
\hline
\multirow{2}{*}{\textbf{HotpotQA}}& PLAID & 26.1 & 7.7\phantom{1} & 56.1\phantom{1}          \\
                                  & \name & \phantom{1}2.5 & 0.08 & 0.09     \\
                                  \hline
\multirow{2}{*}{\textbf{MS MARCO}}& PLAID & 16.5 & 6.7\phantom{1} & 28.1\phantom{1}          \\
                                  & \name & \phantom{1}2.3 & 0.06 & 0.05     \\
\bottomrule

\end{tabular}}
\caption{CPU--GPU data transfer overhead of PLAID and \name across three datasets. 
}
\vspace{-4mm}
\label{tab:transfer}
\end{table}

\begin{figure}[t]
  \centering
  \includegraphics[width=\linewidth]{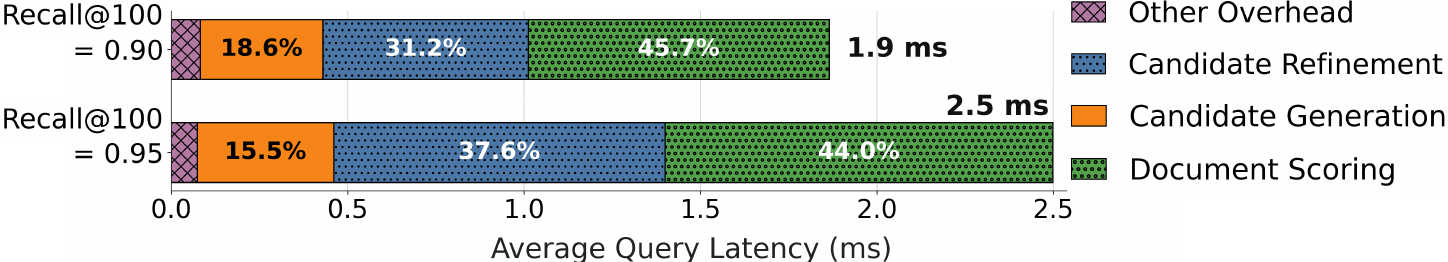}
  \vspace{-7mm}
  \caption{Latency breakdown of \name on LoTTE. 
  }
  \label{fig:lotte_latency_breakdown}
\end{figure}

\stitle{Latency breakdown}
Figure~\ref{fig:lotte_latency_breakdown} shows that \name's latency is dominated by the two compute-intensive stages of the pipeline: candidate refinement and document scoring. On LoTTE, these stages account for over 75\% of the end-to-end latency at both Recall@100 = 0.90 and 0.95, while other overheads---such as memory allocation, data transfer, and runtime bookkeeping---remain small and nearly constant. This suggests that overall latency is primarily driven by multi-vector operations, including distance computation and score aggregation, rather than index probing or system overheads.

We observe similar trends across all datasets and recall targets: increasing the recall target mainly increases the time spent in candidate refinement and document scoring. This observation motivates \name's design, which combines GPU-based fast filtering with chunked execution. The candidate refinement stage efficiently prunes documents using lightweight partial document scoring, thereby reducing the number of candidates passed to the expensive document scoring stage. By further partitioning the workload into chunks, \name overlaps GPU-based 1-bit document scoring with CPU-based full-bit scoring, effectively hiding part of the computation latency and improving end-to-end query throughput.



\rone{\stitle{Index construction time}}
\rone{For offline 4-bit index construction, \name takes 36.4, 40.7, and 73.8 minutes on LoTTE, HotpotQA, and MS MARCO, respectively. This is moderately slower than PLAID, which takes 17.8, 26.6, and 49.8 minutes, but substantially faster than IGP, and GEM, which take 110.5, 105.6, 263.4 minutes and 476, 530, 1016 minutes, respectively. Thus, \name's online performance improvements do not come at the cost of prohibitive offline index construction overhead.}

\stitle{Peak memory consumption}
At Recall@100 = 0.95, \name uses 9.0–17.7 GiB of GPU memory across datasets, peaking at 17.7 GiB on MS MARCO. This shows that \name remains within the memory budget of a single GPU even at large scale.

%% file: sections/related-work.tex
\section{Related Work}
The paradigm of encoding the dataset and queries into multi-vectors and performing retrieval using Chamfer similarity is introduced in ColBERT~\cite{colbert} for document retrieval, which demonstrates significantly better retrieval accuracy than single-vector representations~\cite{luan2021sparse}. Subsequent work has focused on improving the efficiency and scalability of this paradigm. ColBERTv2~\cite{colbertv2} introduces residual compression to significantly reduce the storage overhead of multi-vector representations. Building on this design, PLAID~\cite{plaid} accelerates ColBERTv2 with a two-stage filtering framework that first prunes irrelevant documents using a lightweight centroid interaction mechanism. EMVB~\cite{emvb} further improves upon PLAID by reducing memory consumption and query latency through product quantization~\cite{pq} and SIMD-based optimizations. WARP~\cite{warp} integrates PLAID's techniques with the objective proposed in XTR~\cite{xtr}, avoiding the need to gather complete document representations by computing document scores with missing similarity imputation. DESSERT~\cite{dessert} adopts a different strategy, leveraging a locality-sensitive hashing (LSH)~\cite{lsh}-based randomized indexing scheme to approximate document-query scores via efficient hashing-based estimation. MUVERA~\cite{muvera} takes a different direction by reducing multi-vector retrieval to single-vector maximum inner product search (MIPS), constructing fixed-dimensional encodings whose inner products approximate Chamfer similarity and thus enabling the use of standard single-vector indexes. IGP~\cite{igp} proposes an incremental next-similar retrieval technique over a proximity graph index to support high-quality candidate generation, achieving state-of-the-art performance in multi-vector retrieval. Finally, GEM~\cite{gem} constructs a native set-level proximity graph directly over vector sets, organizing them with TF-IDF-guided clustering and connecting local cluster graphs through shared-set bridges.

%% file: sections/conclusion.tex
\section{Conclusion}
In this paper, we study the problem of multi-vector retrieval for semantic search applications that require fine-grained matching. We focus on GPU-accelerated solutions and observe that CPU--GPU data movement bottlenecks the execution time of existing systems because vector data must be transferred from host memory to the GPU at query time. We propose \name, which relies on hybrid-precision storage---keeping highly compressed quantization codes on the GPU and high-precision data on the CPU---to eliminate this transfer bottleneck. \name carefully coordinates CPU and GPU workloads to maximize resource utilization, which is achieved through a concurrent collaborative scoring scheme that overlaps CPU fine-grained evaluation with GPU coarse filtering. Through extensive experiments on real-world datasets, we show that \name achieves up to $59.5\times$ higher query throughput at the same recall level compared to existing approaches. Remarkably, \name completely removes cross-device data movement as a bottleneck in multi-vector retrieval while fully exploiting the complementary strengths of both processors.